\documentclass[preprint2]{emulateapj}
\usepackage{color,epsfig}
\usepackage{natbib,graphicx,amsmath}
\input{epsf}
\usepackage{epstopdf}
\DeclareGraphicsExtensions{.jpg,.pdf,.png,.eps,.ps}
\graphicspath{{FIGURES/}}
\newcommand{\ltsima}{$\; \buildrel < \over \sim \;$}
\newcommand{\ltsim}{\lower.5ex\hbox{\ltsima}}
\newcommand{\annz}{{\sc ann}{\it z}}
\newcommand{\zphot}{\ensuremath{z_{\mathrm{phot}}}}
\newcommand{\zspec}{\ensuremath{z_{\mathrm{spec}}}}

\shorttitle{BCS Survey}
\shortauthors{Desai et~al.}

\def\Munich{1}
\def\ExcellenceCluster{2}
\def\MPE{4}
\def\Upenn{3}
\def\IAPFrance{5}
\def\Fermilab{6}
\def\UNDakota{7}
\def\UCI{8}
\def\UCOLick{9}
\def\UChicago{10}
\def\Tokyo{11}
\def\IAAASTaiwan{12}
\def\CUTaiwan{13}
\def\STSI{14}
\def\Michigan{15}

\begin{document}
\title{The Blanco Cosmology Survey: Data Acquisition, Processing, Calibration, \\
Quality Diagnostics and Data Release }

\author{
S.~Desai\altaffilmark{\Munich,\ExcellenceCluster},
R.~Armstrong\altaffilmark{\Upenn},
J.J.~Mohr\altaffilmark{\Munich,\ExcellenceCluster,\MPE}, 
D.R.~Semler\altaffilmark{\Munich,\ExcellenceCluster},
J.~Liu\altaffilmark{\Munich,\ExcellenceCluster},
E.~Bertin\altaffilmark{\IAPFrance},
S.S.~Allam\altaffilmark{\Fermilab},
W.A.~Barkhouse\altaffilmark{\UNDakota},
G.~Bazin\altaffilmark{\Munich,\ExcellenceCluster},
E. J.~Buckley-Geer\altaffilmark{\Fermilab},
M.C.~Cooper\altaffilmark{\UCI},
S.M.~Hansen\altaffilmark{\UCOLick},
F.W.~High\altaffilmark{\UChicago},
H.~Lin,\altaffilmark{\Fermilab},
Y.-T.~Lin\altaffilmark{\Tokyo,\IAAASTaiwan},
C.-C.~Ngeow\altaffilmark{\CUTaiwan},
A.~Rest\altaffilmark{\STSI},
J.~Song\altaffilmark{\Michigan},
D.~Tucker\altaffilmark{\Fermilab},
A.~Zenteno\altaffilmark{\Munich,\ExcellenceCluster}
}

\altaffiltext{\Munich}{Department of Physics, Ludwig-Maximilians-Universit\"{a}t, Scheinerstr.\ 1, 81679 M\"{u}nchen, Germany}
\altaffiltext{\ExcellenceCluster}{Excellence Cluster Universe, Boltzmannstr.\ 2, 85748 Garching, Germany}
\altaffiltext{\Upenn}{Department of Physics and Astronomy,University of Pennsylvania, Philadelphia, PA 19104} 
\altaffiltext{\MPE}{Max-Planck-Institut f\"{u}r extraterrestrische Physik, Giessenbachstr.\ 85748 Garching, Germany}
\altaffiltext{\IAPFrance}{Institut d'Astrophysique de Paris, UMR 7095 CNRS, Universit\'e Pierre et Marie Curie, 98 bis boulevard Arago, F-75014 Paris, France}
\altaffiltext{\Fermilab}{Fermi National Accelerator Laboratory, P.O. Box 500, Batavia, IL 60510}
\altaffiltext{\UNDakota}{Department of Physics \& Astrophysics, University of North Dakota, Grand Forks, ND 58202}
\altaffiltext{\UCI}{Hubble Fellow, Department of Physics \& Astronomy, Frederick Reines Hall, University of  California, Irvine, CA 92697, USA} 
\altaffiltext{\UCOLick}{University of California Observatories \& Department of Astronomy, University of California, Santa Cruz,CA 95064}
\altaffiltext{\UChicago}{University of Chicago, 5640 South Ellis Avenue, Chicago, IL 60637}
\altaffiltext{\Tokyo}{Institute for Physics and Mathematics of the Universe, University of Tokyo, 5-1-5 Kashiwa-no-ha, Kashiwa-shi, Chiba 277- 8568, Japan}
\altaffiltext{\IAAASTaiwan}{Institute of Astronomy \& Astrophysics, Academia Sinica, Taipei, Taiwan}
\altaffiltext{\CUTaiwan}{Graduate Institute of Astronomy, National Central University, No. 300 Jhongda Rd, Jhongli City 32001 Taiwan}
\altaffiltext{\STSI}{Space Telescope Science Institute, 3700 San Martin Dr., Baltimore, MD 21218}
\altaffiltext{\Michigan}{Department of Physics, University of Michigan, 450 Church St, Ann Arbor, MI 48109}

\email{shantanu@usm.lmu.de}

\begin{abstract}
The Blanco Cosmology Survey (BCS) is a 60 night imaging survey of $\sim$80~deg$^2$ of the southern sky located in two fields: ($\alpha$,$\delta$)= (5~hr, $-55^{\circ}$) and (23~hr, $-55^{\circ}$). The survey was carried out between 2005 and 2008 in $griz$ bands with the Mosaic2 imager on the Blanco 4m telescope.  The primary aim of the BCS survey is to provide the data required to optically confirm and measure photometric redshifts for Sunyaev-Zel'dovich effect selected galaxy clusters from the South Pole Telescope and the Atacama Cosmology Telescope.  We process and calibrate the BCS data, carrying out PSF corrected model fitting photometry for all detected objects.  The median 10$\sigma$ galaxy (point source) depths over the survey in $griz$ are approximately 23.3 (23.9), 23.4 (24.0), 23.0 (23.6) and 21.3 (22.1), respectively.  The astrometric accuracy relative to the USNO-B survey is $\sim45$~milli-arcsec.  We calibrate our absolute photometry using the stellar locus in $grizJ$ bands, and thus our absolute photometric scale derives from 2MASS which has $\sim2$\% accuracy.  The scatter of stars about the stellar locus indicates a systematics floor in the relative stellar photometric scatter in $griz$ that is $\sim$1.9\%, $\sim$2.2\%, $\sim$2.7\% and$\sim$2.7\%, respectively.  A simple cut in the {\it AstrOmatic} star-galaxy classifier {\tt spread\_model} produces a star sample with good spatial uniformity.  We use the resulting photometric catalogs to calibrate photometric redshifts for the survey and demonstrate scatter $\delta z/(1+z)=0.054$ with an outlier fraction $\eta<5$\% to $z\sim1$.  We highlight some selected science results to date and provide a full description of the released data products.
\end{abstract}


\bigskip\bigskip

\section{Introduction}
\label{sec:intro}
Since the discovery of cosmic acceleration at the end of the last millenium \citep{schmidt98,perlmutter99a}, understanding the underlying causes has remained as one of the key mysteries in modern astrophysics.  As the most massive collapsed structures in the universe, galaxy cluster populations and their evolution with redshift provide a powerful probe of, for example, the dark energy equation of state parameter as well as alternate gravity theories which mimic cosmic acceleration \citep{wang98,haiman01,holder01b}.  Evolution of the cluster abundance depends on a combination of the angular-diameter distance vs. redshift relation and the growth rate of density perturbations.  This sensitivity enables one to constrain a range of cosmological parameters, including the matter density, the sum of the neutrino masses \citep{numass}, the present day amplitude of density fluctuations, and the presence of primordial non-Gaussianity in the initial density fluctuations~\citep{dalal08,cunha10}. 
In addition, galaxy clusters provide an ideal laboratory to study galaxy evolution~\citep[e.g.][]{dressler80}.  Interesting studies of the galaxy properties and their evolution within clusters include studies of the blue fraction and the halo occupation distribution \citep[e.g.][]{butcher84,lin03b,lin04b,lin06,hansen09,zenteno11}.

The first large scale attempt to identify and catalog galaxy clusters was by Abell in 1958. He discovered  galaxy clusters by looking for over-densities of galaxies in Palomar Observatory photographic plates within  a radius of about 2.1~Mpc around a given cluster position~\citep{abell58}. Abell's catalogs  contained about 4700 clusters~\citep{abell89}. However Abell's catalog suffered from incompleteness 
and contamination from projection effects as well as human bias~\citep{Biviano2008}.  With the advent of 
CCD cameras one could apply objective automated algorithms to look for galaxy clusters, and this has led to significant progress in cosmological as well astrophysical studies using galaxy clusters.   

In the last decade, many optical photometric surveys  such as SDSS, CFHTLS, RCS covering contiguous regions of the sky have discovered
several new galaxy clusters spanning a broad range of masses and redshifts. The CFHTLS-W~\citep{Adami2009} has  observed about 171 deg$^2$ in $griz$ bands with  80 \% completeness up 
to  $i$ band  magnitude  of 23.   The RCS-2~\citep{gilbank11} survey  has covered approximately 1000 deg$^2$ in $grz$ bands with 10$\sigma$ magnitude depths of around 23.55 in $r$-band. The SDSS MaxBCG catalog ~\citep{Koester2007}  has covered  about 7,500 deg$^2$ in $ugriz$ bands and with 
10 $\sigma$ $r$ band magnitude  limits of about 22.35.  The largest optical galaxy cluster survey in terms 
of area is the Northern Sky Optical Cluster Survey III which has imaged about 11,400 deg$^2$ up to a redshift of about 0.25~\citep{Gal09}.  The deepest optical cluster survey  to date is the CFHTLS-D survey~\citep{Adami2009}, which reaches 80 \% completeness   for $i$ band magnitudes of  26 and has detected clusters up to redshift of 1.5.  Two upcoming photometric galaxy cluster surveys which will start around October 2012  include DES which will cover about 5,000  deg$^2$  in $grizY$ bands with 10 $\sigma$ $r$-band limiting magnitudes of 24.8,  
and KIDS~\citep{dejong12} which will cover 1,500 deg$^2$ in $ugri$ bands with 10$\sigma$  $r$-band limiting magnitude of 24.45.

One can use such surveys for cosmological studies using galaxy clusters.  For example, ~\citet{gladders07} showed that a large optical galaxy cluster survey could constrain cosmological parameters using the self-calibration method~\citep{majumdar03,majumdar04}.  First cosmological constraints using SDSS optical catalogs are described in ~\citet{rozo10}.  

Over the last decade there have been several mm-wave cluster studies in the Southern Hemisphere, including ACBAR~\citep{acbar}, ACT~\citep{act2004}, APEX~\citep{apex2010} and SPT~\citep{ruhl04}.  All these projects have attempted to carry out galaxy cluster surveys using the Sunyaev-Zel'dovich effect (SZE).  The SZE is the distortion of cosmic microwave background spectrum due to inverse Compton scattering of CMB photons by hot electrons in galaxy clusters~\citep{sunyaev72}, and it provides a promising way to discover galaxy clusters.  Because the surface brightness of the SZE signature of a particular cluster is independent of redshift, SZE survey cluster samples can in principle have sensitivity over a broad range of redshifts~\citep{birkinshaw99,carlstrom02}.  However, to  make use of SZE selected galaxy cluster samples, one needs a well-understood  selection of galaxy clusters (sample contamination and completeness), cluster redshift estimates and a link between SZE signature and the cluster halo masses.  It is important to note that redshift estimates cannot be obtained using SZ experiments alone, and so one needs dedicated optical surveys to follow up these galaxy clusters detected by SZ surveys.

The Blanco Cosmology Survey (BCS) is an optical photometric survey which was designed for
 this purpose and positioned to overlap the ACBAR, ACT, APEX and SPT  surveys in the southern hemisphere.  The goal of BCS is to enable cluster cosmology by providing the data to confirm galaxy clusters from the above surveys and to measure their photometric redshifts. This was done by  surveying two patches totalling $\sim$80~deg$^2$ positioned so that they could be observed with good efficiency over the full night during the period October -- December from Chile. The BCS observing strategy was chosen to obtain depths roughly two magnitudes deeper than SDSS, so that one could estimate photometric redshifts for $L \geq L_{*}$  galaxies out to a redshift $z=1$. 

The outline of this paper is as follows: Sect.~\ref{sec:bcssurvey} describes the Blanco Cosmology Survey, including the camera, observing strategy and site characteristics.  In Sect.~\ref{sec:processing}  we describe in detail the processing and calibration of the dataset using the Dark Energy Survey Data Management System.   In Sect.~\ref{sec:bcsdata},  we describe the photometric characteristics of the BCS dataset and present single galaxy photometric redshifts that are tuned using fields containing large numbers of spectroscopic redshifts.   In this paper all magnitudes refer to
AB magnitudes.


\begin{figure}
\begin{center}
\includegraphics[width=0.49\textwidth]{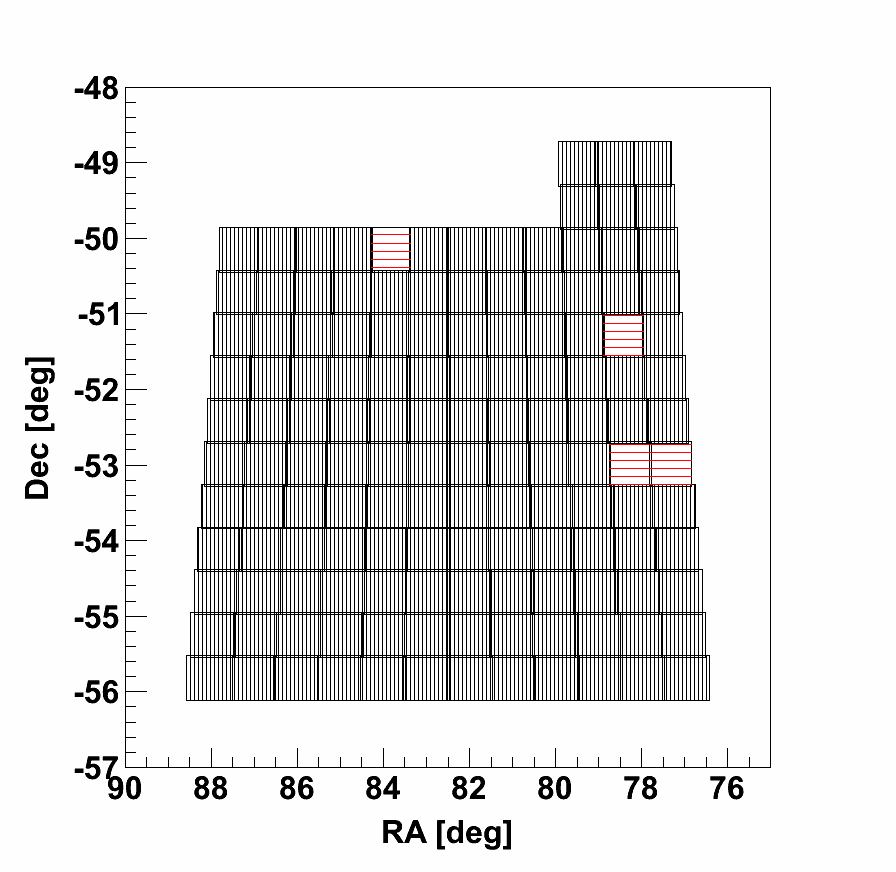}
\vskip-0.10in
\includegraphics[width=0.49\textwidth]{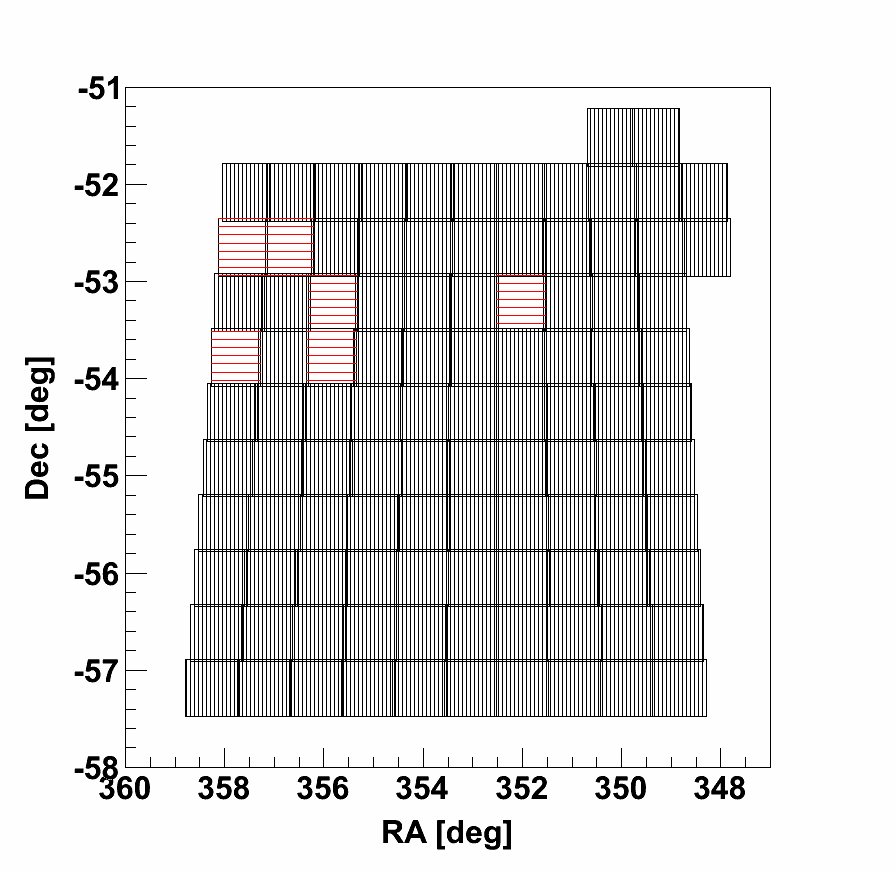}
\caption{BCS survey footprint of coadded tiles in the 5~hr and 23~hr fields. There are 104 tiles covering $\sim$35~deg$^2$ in the 23~hr field and 138 tiles covering $\sim$45~deg$^2$ in the 5~hr field for a total coverage of $\sim$80~deg$^2$.  The black vertically hatched boxes represent tiles which have passed our quality checks. The red horizontally hatched boxes represent tiles with some data quality problems that we have not corrected.}

\label{fig:bcscoverage}
\end{center}
\end{figure}

\section{BCS survey}
\label{sec:bcssurvey}

Blanco Cosmology Survey was a NOAO  Large Survey project (2005B-0043, PI: Joseph 
Mohr) which was awarded 60 nights between 2005 (starting from semester 2005B) and 2008 on the Cerro Tololo Inter American Observatory (CTIO) Blanco 4~m telescope using the Mosaic2 imager with $griz$ bands.  Because of shared nights with other programs, the data acquisition included 69 nights, and the final processed dataset only consists of 66 nights, because two nights were entirely clouded out and the pointing solution for one night (20071105) was wrong due to observer error.  We now describe the Mosaic2 imager on the Blanco telescope and then discuss the BCS observing strategy.

\subsection{Mosaic2 Imager}
The Mosaic2 imager is a prime focus camera on the Blanco 4m telescope that contains eight $2048\times4096$ CCD detectors. The 8 SITe $2\mathrm{K}\times4\mathrm{K}$ CCDs are read out in dual-amplifier
mode, where different halves of each CCD are read out in parallel through separate amplifiers.
The CCDs are read out through a single amplifier per chip simultaneously to 8 controller inputs. 
Read noise is about $6-8$ electrons and readout time is about 110 seconds. The dark current rate is less than 1 electron/pixel/hour at $90$~K. The resulting mosaic array is a square of about 5 inches on an edge. The gaps between CCDs are kept to about 0.7 mm in the row direction and 0.5 mm in the column direction. Given the fast optics at the prime focus on the Blanco, the pixels subtend $0.27''$ on the sky. 
Total field of view is 36.8 arc-minute on a side for a total solid angle per exposure of $\sim$0.4~deg$^2$. More details on the Mosaic2 imager can be found in the online CTIO documentation\footnote{{\tt http://www.ctio.noao.edu/mosaic/manual/index.html}.}.

\subsection{Field selection and multi-wavelength coverage} 
The survey was divided into two fields to allow efficient use of the allocated nights
between October and December. Both fields lie near $\delta = -55^{\circ}$ which allows for overlap with
the SPT and other mm-wave surveys.  One field is centered near $\alpha = 23.5$~hr and the other is at $\alpha$ = 5.5~hr. 
The 5~hr 30~min $-52^{\circ}$ patch consists of a $12\times11$ array of Blanco pointings
 and the 23~hr $-55^{\circ}$ patch is a $10\times10$ array of pointings.
The 5~hr field lies within the Boomerang field where the ACBAR experiment took data. The 23~hr field
has been observed by the APEX, ACT and SPT experiments.
In addition to the large science fields, BCS also covers nine small fields that overlap large spectroscopic
surveys so that photometric redshifts using BCS data can be trained and tested
using a sample of over 5,000 galaxies with spectroscopic redshifts. BCS also surveyed standard star fields for photometric calibration.  The coverage of BCS in 5~hr and 23~hr fields is shown in Figs~\ref{fig:bcscoverage}. For convenience of data processing and building catalogs, we divide the survey region into $36' \times 36'$ square regions called tiles.  Each tile is a $8192\times8192$ pixel portion of a tangent plane projection. These tiles are set on a grid of point separated by $34'$, allowing for approximately $1'$ overlaps of sky between neighboring tiles.  The black vertical hatches in Fig.~\ref{fig:bcscoverage} indicate locations of tiles  which passed various quality checks. The red horizontal hatches indicate locations of tiles which were observed and processed, but failed data quality checks.

We also secured other multi-wavelength observations overlapping parts of the BCS fields.  About 14~deg$^2$ of the 23~hr BCS field was surveyed using {\it XMM-Newton} (known as XMM-BCS survey) and results from those observations are reported elsewhere~\citep{suhada12}.  An $\sim$12~deg$^2$ region of the same field was also targeted in a {\it Spitzer} survey (S-BCS).  More recently, the {\it XMM-Newton} survey has been expanded to 25~deg$^2$, and the {\it Spitzer} survey has been expanded to 100~deg$^2$.  Most of the BCS region has been observed in the near-infrared as part of the the ESO VISTA survey program~\citep{VISTA2011}.

\begin{figure}
\begin{center}
\includegraphics[width=0.49\textwidth]{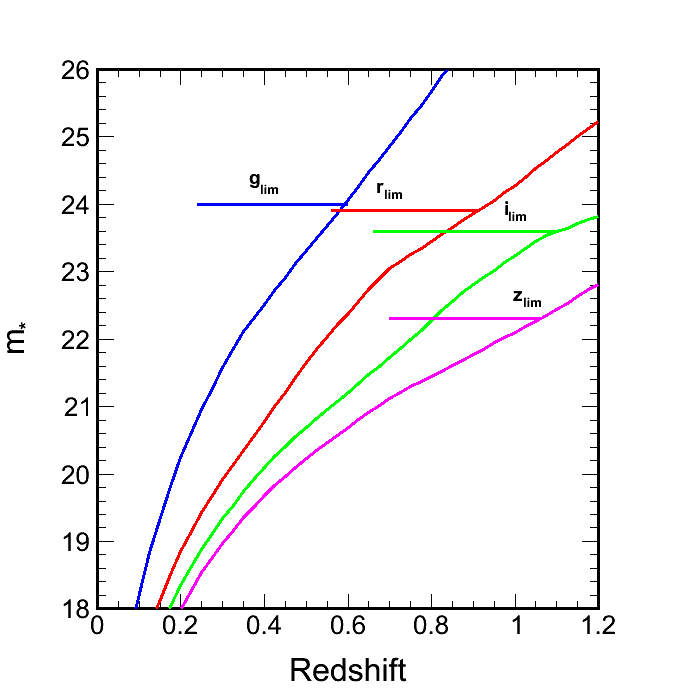}
\caption{Redshift evolution of a passively evolving $L_{*}$ galaxy along with target 10$\sigma$ photometric BCS depths in each band.  The exposure times in each band were tuned so that photometric depth meets or exceeds $L_{*}$ out to the redshift where the 4000\AA break shifts out of that band, but also limited to $z=1$ due to the low sensitivity of the Mosaic2 camera in the $z$-band.}
\label{bcsdepth}
\end{center}
\end{figure}

\subsection {Observing Strategy}
The BCS observing strategy was designed to allow us to accurately measure cluster photometric redshifts out to redshift $z=1$.
Because the 4000\AA\  break is redshifting out to 8000\AA\ by $z=1$, obtaining reliable 
photo-z's for $0< z<1$ requires all four photometric bands $g$, $r$, $i$, $z$ (i.e. one loses all clusters 
at $z<0.4$ if you drop $g$-band and with $z$-band we can actually push beyond $z=1$).   The redshift
 at which the 4000\AA\ break redshifts beyond a particular band sets, crudely speaking, the maximum redshift for which that band is useful for cluster photo-z's; for $griz$ this is $z=0.35$, 0.7, 1.0 and 1.4, respectively.  Because the central wavelength of the g-band is about 4800\AA\, with a FWHM of 1537\AA\, we start losing sensitivity to  very low  redshift clusters, because it is not possible to straddle the 4000\AA\  break. Although detailed studies of the sensitivity of optical cluster detection at low redshifts have not been done, our ability to estimate unbiased red sequence redshifts for clusters is reduced below redshifts $z\sim0.1$ .

We calculate our photometric limits in each band by requiring that the depth allows us to probe at least to $L_*$ at that maximum redshift with 10$\sigma$ photometry.  We use a Bruzual and Charlot $z=3$ single burst model with passive evolution~\citep{bruzual03} to calculate the evolution of $L_*$ in the four bands (see Fig.~\ref{bcsdepth}).  We select our $z$ depth to probe to $L_*$ at $z=1$ rather than at $z=1.4$, because of the low sensitivity of the Mosaic2 detectors in the $z$-band.  The survey was  designed to reach 10$\sigma$ photometric limits within a 2.2 arcsec aperture of $g=24.0$, $r=23.9$, $i=23.6$ and $z=22.3$. These limits assume an airmass of 1.3 and $0.9''$ median seeing for all bands. Assuming bright time for $z$ and $i$ and dark time for $g$ and $r$ these limits require exposures of 250~s, 600~s, 1400~s and 700~s in $griz$, respectively.


In all, we observed about 288 tiles spanning our survey fields. For each field we typically took 2 exposures in $g$ of 125 secs each, 2 exposures in $r$ of 300s each, 3 exposures in $i$ of 450s each and three exposures in $z$ of 235 seconds each.  A limitation of the Mosaic2 detector is a very low saturation of around 25,000~ADU for most of the detectors, and this forced us to take short exposures even through the readout time for each was quite high.  Neighboring pointings have small overlaps, but multiple exposures were offset by approximately half the width of an amplifier to help us tie the survey together photometrically.  Having two shifted exposures allows us to largely overcome the gaps in our survey left by spaces between neighboring chips.   In addition to this primary survey tiling, we also constructed another layer of tilings which was designed to sit at the vertices of unique groups of four adjacent primary pointings. These tiles were observed using shorter exposures during poor seeing conditions on photometric nights.  The 110~s readout of the Mosaic2 camera makes the efficiency of short exposures low, and so in each band we have chosen the minimum number of exposures allowable given the sky brightness. The total exposure per tile is 3000~s and after including the readout time, the total observation time per science field is about 4200~s, giving us an overall efficiency of about 70\%. The dome flats and bias frames were taken in the afternoon, and we did not take any twilight flats.  Over the course of the survey we acquired just over 3000 science exposures and an additional 455 photometric overlap exposures.

In addition to science exposures, on photometric nights we also observed  photometric calibration fields as well as fields for calibrating our photo-z algorithms. These fields were CNOC2, DEEP, CFRS, CDFS, SSA22 and VVDS fields. For the photometric calibration fields we typically observed two or three fields during evening and morning twilight and a single field during the transition from the 23~hr field to the 5~hr~30~min field.  We observed in all four bands during these calibration exposures.  The spectroscopic standard fields were observed to full science depth using the same strategy as for the full survey.

\begin{figure}
\begin{center}
\includegraphics[width=0.49\textwidth]{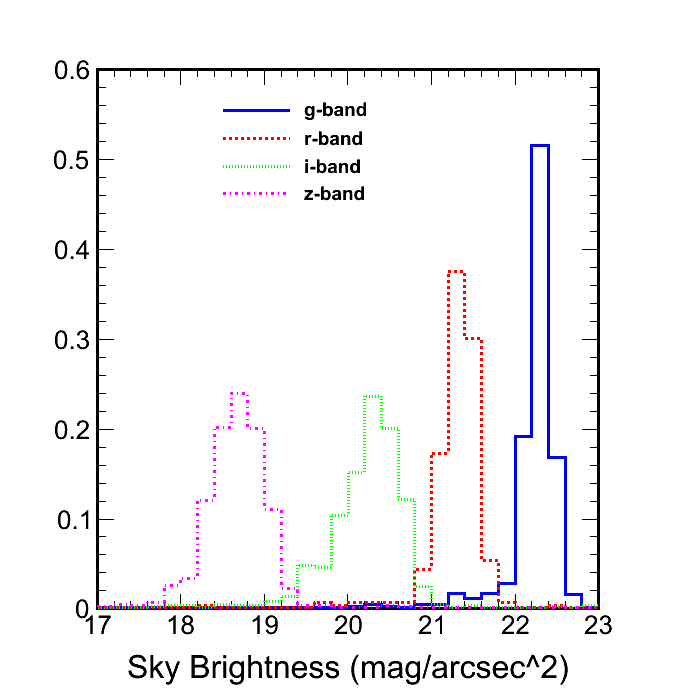}
\caption{Sky brightness distributions for all four bands averaged on  a per exposure basis during the BCS survey.  Typically we observed in $g$ and $r$ during dark time and $i$ and $z$ during bright time. The brightness valuesare peaked at around 22.5, 21.5, 20.5, and 18.75  mag/arcsec$^2$ in $griz$ bands, respectively.}
\label{skybrightness}
\end{center}
\end{figure}
\subsection{Site Characteristics}

The BCS survey provides a sampling of the CTIO site characteristics over a 69 night period in the October to December timeframe over four observing seasons.  Because this is the same timeframe planned for DES observations this provides an interesting glimpse into the expected site characteristics for DES.  Given that the entire Mosaic2 camera and wide field corrector are being replaced by DECam and the new DECam wide field corrector \citep{decam}, the seeing distribution for the DES data could be significantly improved relative to the BCS seeing distribution.

The seeing distribution is shown in top panel of Fig.~\ref{targetpsf}.  The seeing was obtained by running {\tt PSFEX} software on all single-epoch images and using the {\tt FWHM\_MEAN} parameter. The {\tt FWHM\_MEAN} is derived from elliptical Moffat fits to the non-parametric PSF models. These FWHMs include the pixel footprint.  The modal seeing values integrated over the survey are $\simeq 1'$, $0.95'$, $0.8'$, $0.95'$ for $griz$ bands, respectively.  The median seeing values are 1.07, 0.99, 0.95, and 0.95 arcseconds, while the upper and lower quartile seeing values are [0.96, 1.26], [0.89, 1.16], [0.84, 1.13], [0.83, 1.11] arcseconds
respectively. 

The sky brightness is shown in Fig.~\ref{skybrightness}. The sky brightness is calculated using $ZP- 2.5\log{B}$, where $ZP$ is the calculated zeropoint for that image and B is the sky brightness in ADU/arcsec$^{2}$.  The sky brightness distributions in the $griz$ bands
have modal values of  approximately 22.5, 21.5, 20.5 and 18.75~mag/arcsec$^2$, respectively.  Moreover, almost all $i$ and $z$ band data were taken with the moon up, while almost all $g$ and $r$ band data were taken with the moon set. The median values are 22.3, 21.3, 20.3 and 18.7~mag/arcsec$^2$, respectively.

\begin{figure}
\begin{center}
\includegraphics[width=0.49\textwidth]{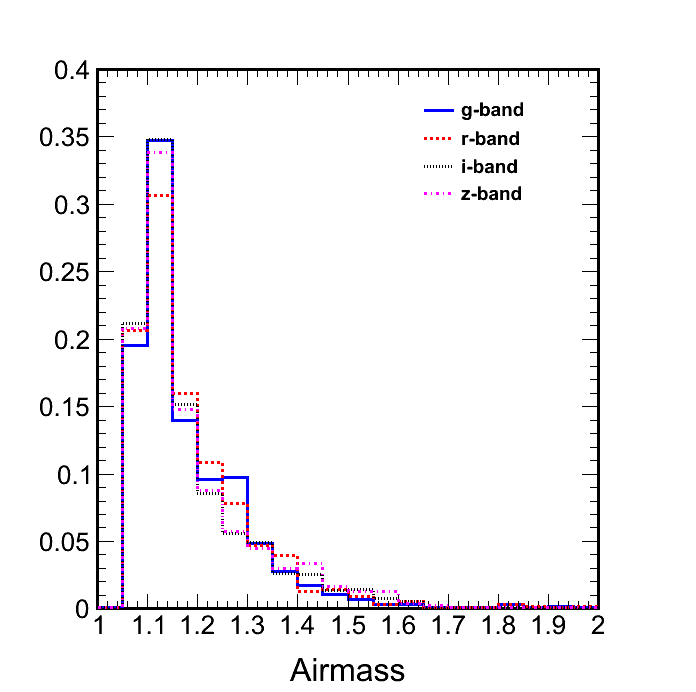}
\caption{The airmass distributions for BCS exposures, color coded by band and normalized by total number of exposures. The peak airmass values in $griz$ bands are 1.144, 1.147, 1.138 and 1.141 respectively.}
\label{airmassfig}
\end{center}
\end{figure}

Given the division of the survey into a 23~hr and a 5~hr field, it was possible to obtain most of the data at relatively low airmass.  
Fig.~\ref{airmassfig} shows the airmass distributions for each band during primary survey observations.  We often obtained photometric calibration field observations over a broader range of airmasses, but we tried to restrict our primary survey observations to air masses of $<$ 1.6.  The median air mass in bands $griz$ are 1.144, 1.147, 1.138 and 1.141 respectively.

\section{Data Processing and Calibration}
\label{sec:processing}

The processing of BCS data is carried out using the automated Dark Energy Survey data management (DESDM) system which has been under development since Fall 2005 at University of Illinois~\citep{ngeow06,mohr08}.  DESDM will be used to process, calibrate and store data from the Dark Energy Survey once it begins operations in October 2012.  Since 2005, the DESDM system has been validated through a series of data challenges with simulated DECam data, which enabled us to improve various steps of the pipeline.  The same automated pipeline was used to analyze BCS data.  The only addition/change to the DESDM pipeline to analyze BCS data was in the crosstalk correction code, for which the routine had to be customized for the Mosaic2 camera.  Processing of the BCS data presented here has been carried out on  National Teragrid resources at NCSA and LONI supercomputers together with dedicated workstations needed for orchestrating the jobs and hosting the database. The middleware for the data reduction pipeline is designed using Condor batch processing system.  Each night takes about 300 CPU hours  for processing.

We have processed BCS data multiple times in a process of discovery where we found problems with the data that required changes to our system.  Scientific results from earlier rounds of processing of BCS data have already appeared, including the optical confirmation of the first ever SZE selected galaxy clusters \citep{staniszewski09} and the discovery of strong gravitational lensing arc using data from the firstå round of processing in Spring 2008~\citep{buckleygeer11}.  Additional galaxy cluster science arising from subsequent rounds of BCS processing have also been published~\citep{high10,zenteno11,suhada12}. Currently our latest processing is being used for additional SZE cluster science within SPT, continued studies of the X-BCS region, and for the followup of the broader XMM-XXL survey over the 23~hr field.

The BCS data were made public one year after their acquisition, as is standard policy at NOAO.  This has enabled multiple independent teams to access the data and use it for their own scientific aims.  The first three seasons of BCS data were processed using an independent pipeline developed at Rutgers University~\citep{menanteau10,menanteau09a}.  All four seasons of BCS data have also been processed independently using NOAO pipeline as part of the current automated processing program, and with the {\tt PHOTPIPE} analysis pipeline~\citep{rest05a}.

\subsection{Detrending}
\label{sec:detrending}
In this section we describe in detail the key steps involved in the DESDM pipeline used for 
reduction of Mosaic2 data to convert raw data products to science ready catalogs and images. Data from every night are processed through a nightly processing or detrending pipeline. Then data from different nights in the same part of the sky are combined using the co-addition pipeline. The detrending pipeline briefly consists of crosstalk corrections, overscan, flatfield, bias and illumination correction, astrometric calibration and cataloging. We now describe in detail each step of the detrending pipeline.

\subsubsection {Crosstalk Corrections}

A common feature of multi-CCD cameras, such as Mosaic2 imager, is crosstalk among the signals from otherwise independent amplifiers or CCDs. This leads to a CCD image containing not only the flux distribution that it collected from the sky, but also a low amplitude version of the sky flux distributions that appear in other CCDs. The crosstalk correction equation is described by:
\begin{equation}
I_{i} = \sum_{j=1}^N \alpha_{ij} I_\mathrm{raw}^{j} ;
\end{equation}
where $I_i$ denotes the crosstalk corrected image pixel value in $i^\mathrm{th}$ CCD, $\alpha_{ij}$ denote the cross-talk coefficients and $I_\mathrm{raw}^{j}$ is the raw image pixel value.  We used cross-talk coefficients provided by NOAO through the survey.  As part of the crosstalk-correction stage, the raw image (which contains 16 extensions) is split into one single-extension file per CCD.  The processing and calibration of CCD mosaics can proceed independently for each CCD after the crosstalk correction, and therefore we split the images to enable efficient staging of the data to the compute resources.

\subsubsection {Image  Detrending}
Detrending is the process that removes the instrumental signatures from the images.  Detrending, in this context, includes overscan correction, bias subtraction, flat fielding, pixel-scale correction, fringe and illumination correction. Both the overscan correction and bias correction are required to remove the bias level present in the CCD and any residual, recurrent structure in the DC bias.  Overscan correction is done for all raw science and calibration images. We subtract the median pixel value in the overscan region in each row for both the amplifiers in each CCD from the raw image pixel values after the crosstalk correction stage.

The median bias frame is created using nightly bias frames taken during the late afternoon, and subtracted from the nightly data. The flat field correction is typically derived from dome flats taken for each observing band. The input dome flat images are overscan corrected, bias corrected, and then scaled to a common mode and then median combined. The resulting flat field correction is scaled by the inverse of the image mode, creating a correction with mean value of about unity. For the bias correction and the flat correction the variation among the input images is used to create an inverse variance weight map that is stored as a second extension in the correction images.  The creation of correction images also requires a bad pixel map, which is an image where pixels with poor response or with high dark current are masked and excluded from the images.  These bad pixel maps are created initially using bias correction and flat field correction images to identify the troublesome pixels.

The bias and flat field corrections are then applied to the science images to remove pixel to pixel sensitivity variations.  These corrections are only applied to those science pixels that are not masked.  In this process, each science image receives an associated inverse variance weight map that encodes the Poisson noise levels and Gaussian propagated noise from each correction step on a per pixel basis.  In addition, each science image has an associated bad pixel map (short integer) where a bit is assigned to each type of masking (i.e. pixels masked from the original bad pixel map, or masked due to saturation, cosmic ray, etc).  In our data model the science image has three extensions:  image, weight map, bad pixel map.  
Each measured flux at the pixel level comes along with its statistical weight and a history of any masking that has been done on that pixel.

For the Mosaic2 imager which has significant focal plane distortion, the pixel scale varies significantly over the field, leading to a significant trend in delivered pixel brightness as a function of position even with a flat input sky.  For such detectors flattening the sky introduces a photometric non-flatness to the focal plane.  Typically this pixel scale variation is corrected during the process of remapping to a portion of a tangent plane, but in our case we prefer to do the single epoch cataloging on images that do not suffer from correlated noise.  Therefore, we  apply a pixel scale correction to account for variation of pixel response as a function of $x$ and $y$ position for each CCD.  We first created master template images to determine photometric flatness corrections using astrometrically refined images from the Mosaic2 camera that we use to calculate the solid angle of each pixel. The correction image is then normalized by the median value, providing a flat field like correction image that can be used to bring all pixels to a uniform flux sensitivity.  To avoid reintroducing trends in the sky with this correction, we apply this correction only to the values of each pixel after subtraction of the modal sky value.  Effectively, this correction scales only source flux while maintaining a flat sky.
  
Illumination and fringe corrections are derived from fully processed science observations in a particular band.  These can be from a single night or shared across nights.  Usually if there was only one exposure from a given band in a night we use science observations from neighbouring nights to create the illumination and  fringe correction images.  Illumination corrections are done for all images, but fringe corrections on the Mosaic2 camera are needed only for $i$ and $z$ bands.  To create these correction images, we first create sky flat templates.  This requires a process of stacking all the detrended images in a band-CCD combination after first flagging all pixels contaminated by source flux.  Source contaminated
pixels are determined by applying a simple threshold above background with a variable grow radius so that all neighboring pixels of a pixel determined to contain source flux are also masked.   Modal sky values are then calculated for each image using pixels that are not flagged for any reason (object pixels, hot column, saturated, interpolated, etc).  The reduced images are then scaled to a common modal sky value, median combined and then rescaled to a unit modal value. 

This science sky image then contains a combination of any illumination and fringe signatures that are common to the input images.  To create the illumination correction we adaptively smooth the science sky images with a kernel that is large in the center and grows smaller near the edges.  This effectively averages out the effects of any fringing, leaving an illumination correction image behind.  The fringe correction is then produced by first differencing the science sky image and the illumination correction image, leaving behind an image of the small scale structure (i.e. fringe signature) that is common to all the science images.  This fringe correction image is then scaled by the model value of the science flat image to produce a fractional fringe correction image.

The illumination correction image is applied like a flat field correction to all previously corrected images, thereby removing any trends that are introduced by the differences in illumination of the dome flats and the flat sky.  The fringe correction is applied by first scaling the correction image by the modal value of the sky in the science image and then subtracting it.  The results of these two corrections are visually very impressive.  The fringe effects in $i$ and $z$ band are nicely removed in almost all cases.  We have found some problem images where the fringe correction leaves clearly visible fringe signatures, and these are cases where only a few frames in $i$ or $z$ were taken on a particular night, and the use of images from neighboring nights to create the corrections was not adequate. 

We expect that the residual scatter we measure could be further reduced using a star flat technique to better characterize the non-uniformities in the pupil ghost.  Nevertheless, the delivered data quality from our current flattening prescription produces data that meet our data quality requirements.  We note that the same prescription has been used previously to meet the data quality requirements of the SuperMACHO experiment in the processing of Mosaic2 data.

At the end of this series of  image detrending steps  which includes overscan, bias, flat-field, pixel-scale, illumination and fringe corrections, the pipeline creates  eight  images (one for each CCD) for every science exposure.  These single epoch image FITS files are called {\it red} images, and they contain 3 extensions:  the main image, a bad pixel mask (BPM)  and an inverse variance weight image. The BPM contains a short integer image where any unusable pixels have non-zero values (coded according to the source of the problem). The weight-map is an inverse variance image map that tracks the noise on the pixel scale and where the weight is set to zero for all masked pixels.
  
\begin{figure}
\begin{center}
\includegraphics[width=0.49\textwidth]{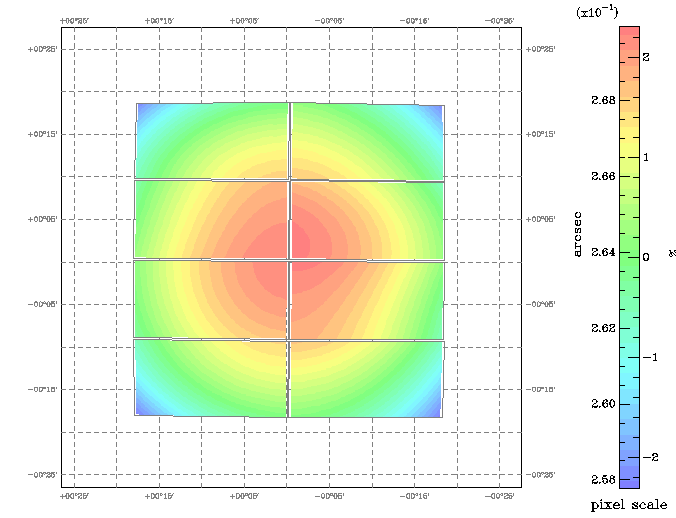}
\caption{Distortion map produced by {\sc SCAMP} for one Mosaic2 exposure consisting of 8 images. TPV distortion model was used. The Mosaic2 distortions were modelled for each CCD by expressing distortions along the RA and DEC direction each with a third order polynomial in CCD $x$ and $y$.}
\label{scampdistortion}
\end{center}
\end{figure}

\subsubsection  {Astrometric Calibration}
Besides pointing errors, wide-field imagers exhibit instrumental distortions that generally deviate significantly from those of a pure tangential projection.  In addition, the vertical gradient of atmospheric refractivity creates a small image flattening of the order of a few hundredths of a percent (corresponding to a few pixels on a large mosaic), with direction and amplitude depending on the direction of the pointing. These three contributions are modeled in the {\sc SCAMP}~\citep{bertin06} package that we use for astrometric calibration.  {\sc SCAMP} uses the TPV distortion model\footnote{currently under review
for inclusion in the registry of FITS conventions; see {\tt http://fits.gsfc.nasa.gov/registry/tpvwcs.html}.}, which maps detector coordinates to true tangent plane coordinates using a polynomial expansion. 

{\sc SCAMP} is normally meant to be run on a large set of {\sc SExtractor} catalogs extracted from overlapping exposures  together with a reference catalog, in order to derive a global solution. 
However, since our pipeline operates on an image-by-image basis, we proceed in two steps: we first run {\sc SCAMP} once on a small subset of catalogs extracted from BCS mosaic images to derive an accurate
 polynomial model of the distortions where the distortions in RA/DEC tangential plane are expressed as a third degree
 polynomial function of the CCD $x$/$y$ position. This Mosaic2 distortion map modeled using a third order polynomial per CCD for a BCS exposure is shown in Fig.~\ref{scampdistortion}.   The astrometric solution computed in this first step of calibration is based on a set of overlapping catalogs from dithered exposures which provides tighter constraints on non-linear distortion terms (than catalogs taken individually).  
Using this model, we create a distortion catalog that encodes the fixed distortion pattern of the detector.  
We then run {\sc SCAMP} on catalogs from each individual exposure (i.e. the union of the  catalogs from each of the eight single epoch detrended images), allowing only linear terms (two for small position offsets and four for the linear distortion matrix) describing the whole focal plane to vary from exposure to exposure.  The solutions for the World Coordinate System (WCS) including the TPV model parameters are then inserted back into image headers.  This approach capitalizes on the expected constancy of the instrumental distortions over time.

\begin{figure}
\begin{center}
\includegraphics[width=0.49\textwidth]{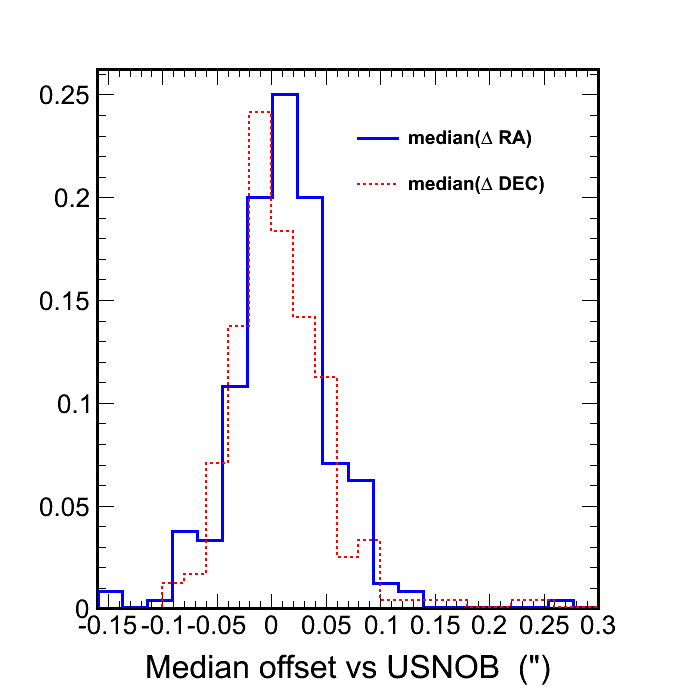}
\caption{Median value of the difference in RA and DEC for objects in BCS coadd catalogs vs USNOB catalog  
for every tile in arcsecs. The matching is done in a $2''$ window. The histograms are peaked at $\sim0.0104''$ and $0.0084''$ in $\Delta$-RA and $\Delta$-DEC respectively. The rms of the histograms in RA and DEC is about 0.047 and 0.045 arcsecs. Note that the intrinsic accuracy of the USNOB catalog is about 0.2 arcsecs~\citep{monet03}.}
\label{scampqa}
\end{center}
\end{figure}

We use the USNO-B1~\citep{monet03} catalog as the astrometric reference.  For astrometric refinement, the cataloging is done using {\sc SExtractor}, and using WINdowed barycenters to estimate the positions of sources.

The astrometric accuracy is quite good, as can be demonstrated with the BCS coadds.  First, the accuracy is at the level of a fraction of a PSF or else significant PSF distortions would appear in the coadds, and this is not the case.  Second, we can measure the absolute accuracy relative to the calibrating catalog USNOB by probing for systematic offsets in RA or DEC between our object catalogs and those from the calibration source.  Fig.~\ref{scampqa} shows the distribution of median offsets within all the coadd tiles for both RA and DEC.  The mean of the histograms is $0.0104''$ in RA and $0.0084''$ in DEC, and the corresponding rms scatter is 47~milli-arcsec and 45~milli-arcsec, respectively.  The USNOB catalog itself has an absolute accuracy with  characteristic uncertainty of 200 milli-arcseconds~\citep{monet03}, which then clearly dominates the astrometric uncertainty of our final catalogs.

\subsubsection {Single-Epoch Cataloging}  
\begin{figure}
\begin{center}
\includegraphics[width=0.49\textwidth]{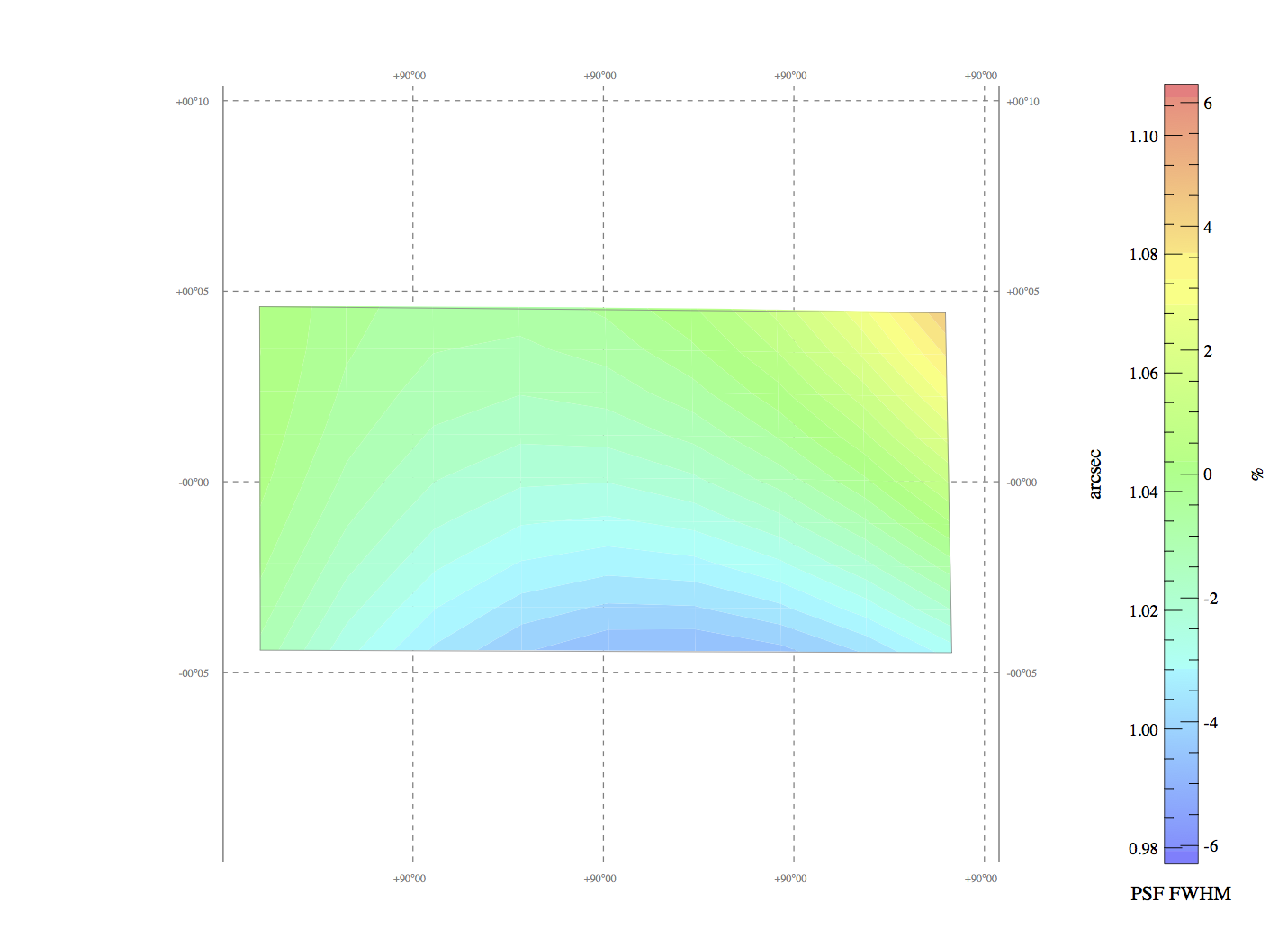}
\caption{Variation of the PSF model FWHM for $g$-band across a single-epoch image from the BCS night 20061030.  Variations across the roughly $10'\times20'$ image are at the 10\% level.}
\label{fig:psfvariationsingle}
\end{center}
\end{figure}

To catalog all objects from single-epoch images we run {\sc SExtractor} using PSF modeling and model-fitting photometry. A PSF model is derived for each CCD image using the {\sc PSFEx} package \citep{bertin11}.  PSF variations within the each CCD are modeled as a $N^\mathrm{th}$ degree polynomial expansion in CCD coordinates.  For our application we adopt a $26\times26$ pixel kernel and follow variations to 3$^\mathrm{rd}$ order.  An example of variation of the Full-Width at Half Maximum (FWHM) of the PSF model across a single-epoch image  is shown in Fig.~\ref{fig:psfvariationsingle}.  The FWHM varies at the 10\% level across this CCD due to both instrumental and integrated atmospheric effects.

\begin{deluxetable}{lr}
\tighten
\tabletypesize{\scriptsize}
\tablecaption{Sextractor Detection Parameters}
\tablewidth{0pt}
\tablehead{
\colhead{Parameter} & \colhead {Values}
}
{\tt DETECT\underline{~}TYPE     }       & {\tt CCD             }    \\
{\tt DETECT\underline{~}MINAREA  }   & {\tt 5               }  \\
{\tt DETECT\underline{~}THRESH   }    & {\tt 1.5             }  \\
{\tt ANALYSIS\underline{~}THRESH }   & {\tt 1.5             }  \\
{\tt FILTER                      }                    & {\tt Y               }  \\
{\tt FILTER\underline{~}NAME     }        & {\tt gauss\underline{~}3.0\underline{~}3x3.conv }   \\
{\tt DEBLEND\underline{~}NTHRESH }  & {\tt 32              }  \\
{\tt DEBLEND\underline{~}MINCONT } & {\tt 0.005           } \\
{\tt CLEAN                       }                   & {\tt Y               }  \\
{\tt CLEAN\underline{~}PARAM     }      & {\tt 1.0             }  \\
{\tt BACKPHOTO\underline{~}THICK }  & {\tt 24.0 } 
\enddata
\label{tab:sex}
\end{deluxetable}

A new version of  {\sc SExtractor} (version 2.14.2) uses this PSF model to carry out PSF corrected model fitting photometry over each image.  The code proceeds by fitting a PSF model and a galaxy model to every source in the image.  The two-dimensional modeling uses a weighted $\chi^2$ that captures  the goodness of fit between the observed flux distribution and the model and iterates to a minimum.  The resulting model parameters are stored and ``asymptotic'' magnitude estimates are extracted by integrating over these models.  This code has been extensively tested within the DESDM program on simulated images, but the BCS data provide the first large scale real world test.  For the BCS application we adopt a S\'ersic profile galaxy model that has an ellipticity and orientation.  This model fitting is computationally intensive and slows the ``lightning-fast'' {\sc SExtractor} down to a rate on the order of 10~objects/s on a single core.  The {\sc SExtractor} config file detection parameters are shown in Table~\ref{tab:sex}.

The advantages of model fitting photometry on single epoch images that have not been remapped are manifold.  First, pixel to pixel noise correlations are not present in the data and do not have to be corrected for in estimating measurement uncertainties.  Second, unbiased PSF and galaxy model fitting photometry is available across the image, allowing one to go beyond an approximate aperture correction to aperture magnitudes often used to extract galaxy and stellar photometry.  Third, there are morphological parameters that can be extracted after directly accounting for the local PSF, which allows for improvements in star-galaxy classification and the extraction of PSF corrected galaxy shear.  A more detailed description of these new {\sc SExtractor} capabilities along with the results from an extensive testing program within DESDM will appear elsewhere (Bertin et al, in preparation).

\subsubsection{Remapping} 
From the WCS parameters which are computed for every reduced image, one can approximate the footprint of the CCD on the sky using frame boundaries in Right Ascension and Declination.  For the BCS survey we have a predefined grid of $36'\times36'$ tangent plane tiles covering the observed fields. Based on this, for every {\it red } image which is astrometrically calibrated, we determine which tiles it overlaps.  We then use {\sc SWarp} \citep{bertin02} to produce background-subtracted remapped images that conform to sections of these tangent plane tiles.  A particular {\it red} image can be remapped to up to four different {\it remap} images in this process.  Pixels are resampled using Lancz\'os-3 interpolation.

Remapping also produces a pixel weightmap and we also remap the bad pixel map (using nearest neighbor remapping). In this process of remapping, zero weight pixels in the reduced images generically impact multiple pixels in the remap image given the size of the interpolation kernel.  These remaps are then stored for later photometric calibration and coaddition.  This on-the-fly remapping need not be done, because at a later stage of coaddition one could in principle return to the {\it red} images, but given the PSF homogenization we do prior to coaddition we have found it convenient to do the remapping as we are processing the nightly datasets.

\subsubsection{Nightly Photometric Calibration}
\label{sec:nightlycalibration}
Our initial strategy for photometric calibration involved traditional photometric calibration using the standard fields observed on photometric nights along with the image overlaps to create a common zeropoint across all our tiles.  In fact, within DESDM we have developed a so-called Photometric Standards Module~\citep{Tucker2007} (PSM)  that we use to fit for nightly photometric solutions, and then we apply those solutions to all science images and associated catalogs from that night.  For BCS 
this involves determining the zeropoints of all images on photometric nights through calibration to identified non-variable Standard stars from SDSS Stripe 82 field~\citep{Smith2002}.


This procedure was used for processing and calibration of the BCS data processed in Spring 2008.  But closer analysis of these data showed that we were not able to control photometric zeropoints to the required level to allow for cluster photometric redshifts over the full survey area.  We therefore abandoned this method for BCS in favor of relative photometric calibration using common stars in overlapping {\it red} images followed by absolute calibration using the stellar locus (described in more detail in Sec.~\ref{sec:absolutephotometry}).  One problem we faced is that so-called photometric nights exhibited non-photometric behavior in the standard field observations.  There was no reliable photometric monitor camera at CTIO during our survey, and so observers simply used the time honored tradition of watching for clouds to make the call on a night being photometric.  Because of our strategy for standard star observations (beginning, middle, end of night), even those nights that exhibit consistent photometric solutions need not have been photometric over the full nights.  Therefore, we felt it safer to assume that no night was truly photometric and to calibrate the data using an entirely different approach.

The results from the PSM module for those nights exhibiting good photometric solutions are still useful.  They have allowed us to monitor changes in the detectors and to measure the color terms in transforming our photometry onto the SDSS system.  We provide a brief description of this procedure, although no science results in this paper are based on PSM related direct photometric calibration.  We expect to apply this method for absolute photometric calibration of DES data where we will indeed have an IR photometric monitoring camera on the mountain.  The PSM solves for the following equation :
\begin{equation}
m_\mathrm{inst} - m_\mathrm{std} = a_n + b_n \times (stdColor - stdcolor_0) + kX
\end{equation}
where $a_n$ is the photometric zeropoint for all 8 CCDs, $b_n$ is the color term, $stdColor$ is the fiducial color around which we define our standard solutions,  which is
$g-r$ for $g$ and $r$ bands, and $r-i$ for $i$ and $z$ bands, $stdcolor_0$ is a constant equal to $g-r=0.53$ for $g$ and $r$ bands and $r-i=0.09$ for $i$ and $z$ bands, $k$ is the first-order extinction coefficient and $X$ is the airmass. The PSM module solves for $a_n$, $b_n$ and $k$  for each photometric night. Using these values for the PSM $a$, $b$ and $k$, one can also estimate the expected zeropoint for every exposure. We calculate it as follows  
\begin{equation}
ZP = -a + 2.5 \log\mathrm{(exptime)} -kX
\end{equation}
We applied the PSM on about 30 BCS nights which  were classified as photometric.  
We also checked for trends in variation of color terms as a function of CCD number. Only the $i$-band color term shows some variation, and this approximate constancy of color terms greatly simplifies the coaddition of the data, because we don't have to track which CCDs have contributed to each pixel on the sky. The color terms we have used for photometric calibration are -0.1221, -0.0123, -0.1907, and 0.0226 in $griz$ respectively.  We also examine the band dependent extinction coefficient ($k$) calculated using data from the photometric nights.  For the ensemble of about 30 photometric solutions in each band, we find the median $griz$ extinction coefficients at CTIO over the life of the survey to be 0.181, 0.104, 0.087 and 0.067 mag./airmass respectively.

This completes the description of all the steps of the nightly processing or single-epoch processing that we do for BCS.  

\subsection{Coaddition}
\label{sec:coaddpipeline}
Once we have data processed for all of the  BCS nights, we then combine data within common locations on the sky to build deeper images that we call coadds. This process is called co-addition and is complicated because it involves combining data taken in widely separated times and under very different observing conditions. Co-addition processing is done on a tile by tile basis. We describe our approach below.

\subsubsection{Relative Photometric Calibration} 
\label{sec:relativecalibration}
During single-epoch processing we extract instrumental magnitudes.  To produce science ready catalogs, we must calculate the zeropoint for every image and re-calibrate the magnitudes. The photometric calibration is done in two steps. The first step is a relative zeropoint calibration that uses the same object in overlapping exposures, and the second is an absolute calibration using the stellar locus.

The relative calibration is done tile by tile rather than simultaneously across the full survey.  We use two different pieces of information to calculate the relative zeropoints.  The primary constraint comes from the average magnitude differences from pairs of {\it red} images with overlapping stars.  The stars are selected based on the {\sc SExtractor} flags and {\tt spread\_model} (discussed later in Sec.~\ref{sec:cataloging})  values. In cases where there aren't enough overlapping stars, we use the average CCD to CCD zeropoint differences derived from photometric nights.  In previous versions of the reduction we also used direct zeropoints derived from photometric nights (see Sec.~\ref{sec:nightlycalibration}) and relative sky brightnesses on pairs of CCDs.  As previously mentioned, the direct photometric zeropoint information is contaminated at some level.  The sky brightness constraints also seem to be problematic for BCS, because only $g$ and $r$ band data were taken on dark nights with no moon present which can introduce a gradient across the camera.  To avoid a degradation of the calibration we used neither the sky brightness constraints nor the direct photometric zeropoints. 

We determine the zeropoints for all images in a tile by doing a least squares solution using the inputs described above.  For this least squares solution there are $N$ input images, each with an unknown zeropoint in the vector $z$. We arbitrarily fix the zeropoint for one image and calibrate the remaining images relative to it. We have $M$ different constraints in the constraint vector $c$. The matrix $A$ is $N \times M$ and denotes the images involved for each constraint. The resulting system of equations is described by $Az = c$ where we use singular value decomposition to solve for the vector $z$.  This gives the relative zeropoints needed to coadd the data for a particular tile.

\subsubsection{PSF Homogenization}
Combining images with variable seeing generically leads to a PSF that varies discontinuously over the coadded image. This affects star galaxy separation and contributes to variation across the image in the completeness at a given photometric depth.  The PSF accuracy could be quite poor in regions where there are abrupt changes to the PSF which would translate into biases in the photometry that would be difficult to track.  The main steps involved in the process of PSF homogenization include: (1) modeling the PSF using {\sc PSFEx} for all {\it remap} images contributing to a coadd tile, (2) choosing the parameters of the target PSF, (3) using {\sc PSFEx} to generate the homogenization kernel, and (4) carrying out the convolution to homogenize all the {\it remap} images to a common PSF. 

\begin{figure}
\begin{center}
\includegraphics[width=0.49\textwidth]{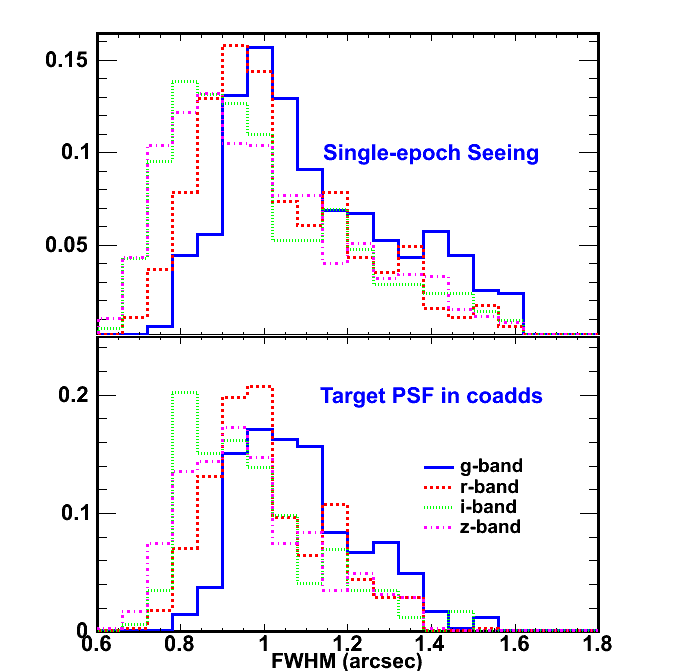}
\caption{FWHM of single epoch images using {\tt PSFex} (top panel) along with the  target PSF FWHM used for homogenizing 
the coadd images for the full BCS survey (bottom panel). The peak values of target PSFs are about $1''$ for $g$ and $r$ bands, $0.9''$ for $i$-band and $0.8''$ for $z$-band respectively.}
\label{targetpsf}
\end{center}
\end{figure}

To reduce PSF variation we  processed our images to bring them to a common PSF within an image and from image to image within a coadd tile.  To do this we apply position dependent convolution kernels that are  determined using power spectrum weighting functions that adjust the relative contributions of large scale and small scale power within an image in such a way as to bring the PSFs within and among the image samples into agreement. The target PSF is defined to be a circular Moffat function with the FWHM set to be the same as the median value of all input PSFs.  

\begin{equation}
\chi^2 = \lvert \Psi - \sum_{l}Y_l(x_i)\kappa_l\ast\Psi_\mathrm{median} \rvert^2
\end{equation}
where $Y_l$ are the elements of a polynomial basis in $x-y$. The target PSF is defined to be a circular Moffat function with the FWHM set to be the median FWHM of the input images.  We imposed a cut on input image PSF $\mathrm{FWHM}<1.6$~arcsec.  This selects only images with relatively good seeing. Images from each band are homogenized separately. The FWHM of the target PSF for all BCS tiles is shown in bottom panel of  Fig.~\ref{targetpsf}.  

Another price of homogenization is that noise is correlated on the scale of the PSF.  While the noise is already correlated to some degree through the remapping interpolation kernel, PSF homogenization characteristically affects larger angular scales than does the remapping kernel.  This leads to biases in photometric and morphological uncertainties, and can also affect initial object detection process in {\sc SExtractor}.  To address this within DES we account for the noise correlations on two critical scales by producing two different weight maps.  The first weight map is used to track the pixel scale noise, and the second weight map is used to correct for the correlated noise on the scale of the PSF.  The pixel scale weight map is used by {\sc SExtractor} in determining photometric and morphological uncertainties.  The PSF scale weight map is used by {\sc SExtractor} in the detection process.  Extensive tests within DESDM have shown this approach to be adequate to produce unbiased photometric and morphological uncertainties and to enable unbiased detection of objects within coadds built from homogenized images.  These results will be presented in detail elsewhere.  For the BCS processing we used only a single pixel-scale weight map, tuned to return the correct measurement uncertainties within {\sc SExtractor}.

\begin{figure}
\begin{center}
\includegraphics[width=0.49\textwidth]{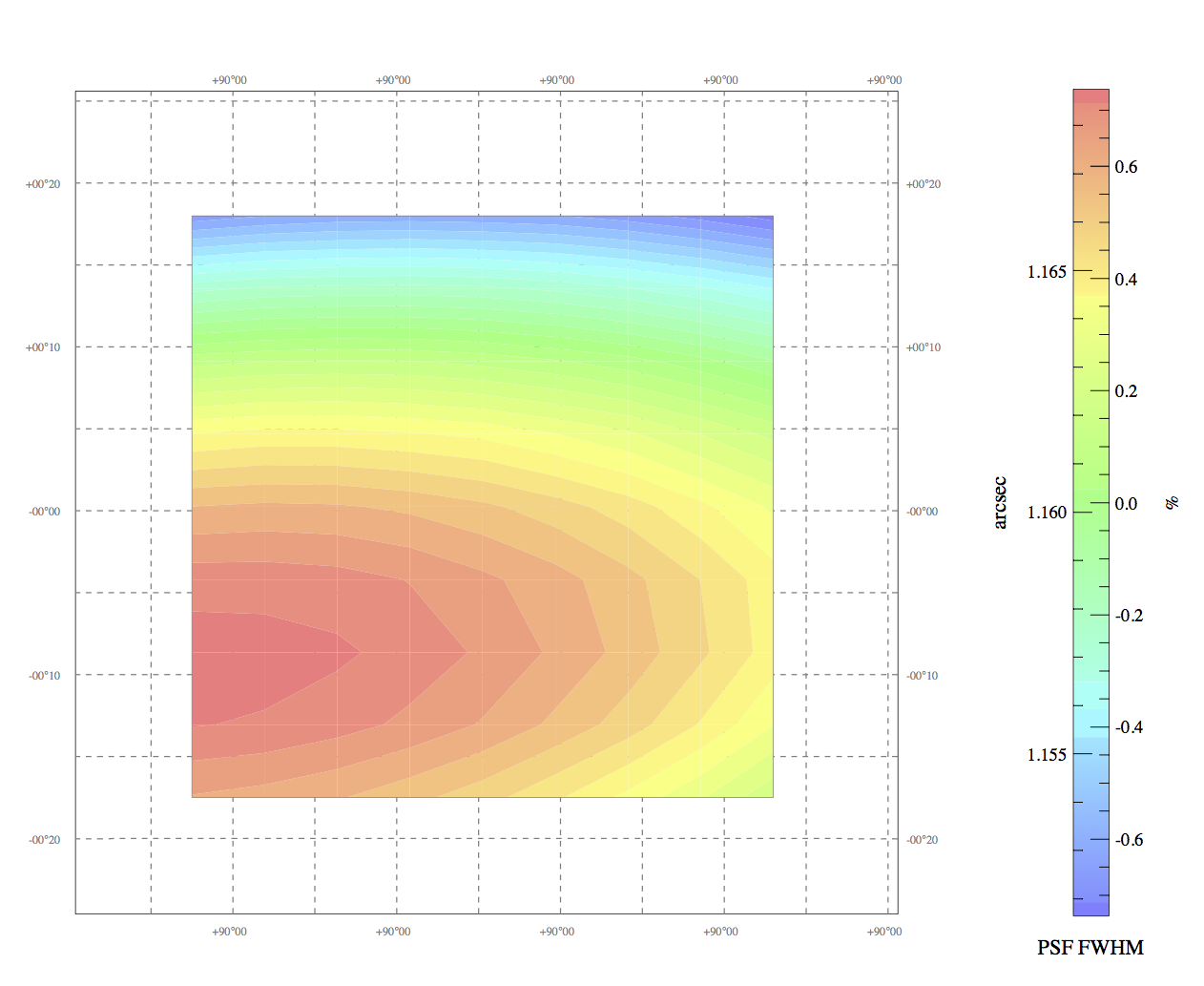}
\caption{Variation of the PSF model FWHM for a $g$-band coadd image for the coadd tile BCS0516-5441.  Because of the homogenization process the variations are at the level of 1\% across the $36'$ image.}
\label{fig:psfvariation}
\end{center}
\end{figure}

\subsubsection {Stacking Single Epoch Images}
\label{coadd}
We use {\sc SWarp} to combine the PSF-homogenized images to build the coadd tile.  Inputs include the relative flux scales derived from the calibration described in Sec.~\ref{sec:relativecalibration}).  We combine the homogenized remap images using the associated weight maps and bpm for each image. The values of the flux-scaled, resampled pixels for each image are then median combined to create the output image. This allows us to be more robust to transient features such as cosmic rays in the $i$ and $z$ bands where there are three overlapping images.  Also, objects with saturated pixels in all single epoch images will contain pixels that are marked as saturated in the coadd images as well.  This ensures accurate flagging of objects with untrustworthy photometry during the coadd cataloging stage. The resulting output coadd image's size is $8192\times 8192$ pixels or approximately $0.6 \times 0.6$ degrees. 

\begin{figure}
\begin{center}
\includegraphics[width=0.49\textwidth]{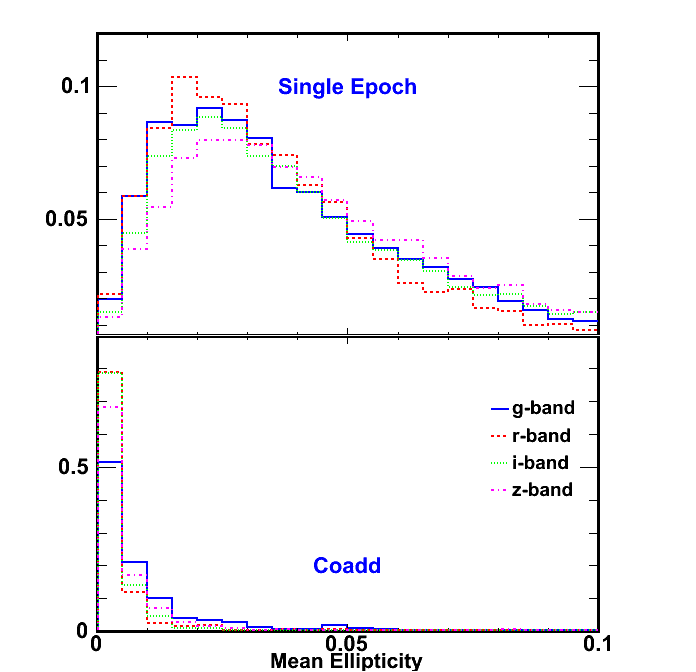}
\caption{Mean Ellipticity calculated by {\sc PSFEx} for single-epoch images (top panel) and for PSF homogenized coadds (bottom panel), color coded by band.  Ellipticity is defined as ${(a-b)}/{(a+b)}$ where $a$ and $b$ refer to semi-major and semi-minor axis respectively. For single-epoch images the median ellipticty for $griz$ bands is  0.0342, 0.0326, 0.0374 and 0.04 respectively. For coadds, typical values are around 0.004, 0.0024, 0.0026, 0.0033.}
\label{fig:ellipticity}
\end{center}
\end{figure}

Fig.~\ref{fig:psfvariation} shows a map of the FWHM as a function of position over one homogenized coadd image.  Variations are at the level of $\sim$1\% over the coadd, as compared to the $\sim$10\% variations that are typical for Mosaic2 across a single CCD (see Fig.~\ref{fig:psfvariationsingle}).  The constancy of PSF as a function of a position ensures that the PSF model can be modeled accurately and that the PSF corrected model fitting photometry is unbiased.  The PSF homogenization process also circularizes the PSF.  Fig~\ref{fig:ellipticity} shows the distribution of ellipticities for the Mosaic2 single epoch images and coadded images~(color coded by band).  The single epoch ellipticity varies up to 0.1 with a modal value around 0.02.  By contrast, the ellipticity distribution of the BCS coadds is peaked at a fraction of a percent with a median value of 0.001.

\subsubsection {Cataloging of Coadded images}
\label{sec:cataloging}
To catalog the objects from coadded images, we run {\sc SExtractor} in dual-image mode with a common detection image across all bands.  For BCS we use the $i$-band  image as the detection image, because it has three overlapping images so the cosmic ray removal is good, and it is by design the deepest of the bands.  We then run {\sc SExtractor} using model-fitting photometry using this detection image
and coadded image in each band. This ensures that a common set of objects are cataloged across all bands.  In both single epoch and coaddition cataloging, the detection criterion was that a minimum of 5 adjacent pixels had to have flux levels about 1.5$\sigma$ above background noise.  The full {\sc SExtractor} detection parameters used for both coadded and single-epoch images are shown in Tab.~\ref{tab:sex}.  In all we catalog about 800 columns across four bands. However for the public data release, we have released 60 columns from {\sc SExtractor}  per object.  This full list in can be found in Tab.~\ref{tab:bcscatalog}.  Most of the parameters are described in the {\sc SExtractor} manual online.  
There are a few additional parameters which are not yet released in the public version of {\sc SExtractor}.  
These include model magnitudes and a new star-galaxy classifier called {\tt spread\_model}, which is a normalized simplified linear discriminant between the best fitting  local PSF model ($\phi$)  and a slightly more extended  model ($G$)  made from the same PSF convolved  with a circular exponential disk model with $\mathrm{scale-length} = \mathrm{FWHM}/16$ 
 (where $FWHM$ is the full-width-half maximum  of the PSF model).  It can be defined by the following equation 
\begin{equation}
{\tt spread\_model} = \frac{{\bf \phi^T x}}{{\bf \phi^T\phi}} - \frac{{\bf G^T x}}{{\bf G^T\phi}} ; 
\end{equation}
where $x$ is the image vector centered on the source. The distribution of {\tt spread\_model} for BCS catalogs is discussed in ~\ref{subsec:sgseparation}.  
More details of {\tt spread\_model} will be described elsewhere (Bertin et al in preparation).

\begin{figure}
\begin{center}
\includegraphics[width=0.40\textwidth]{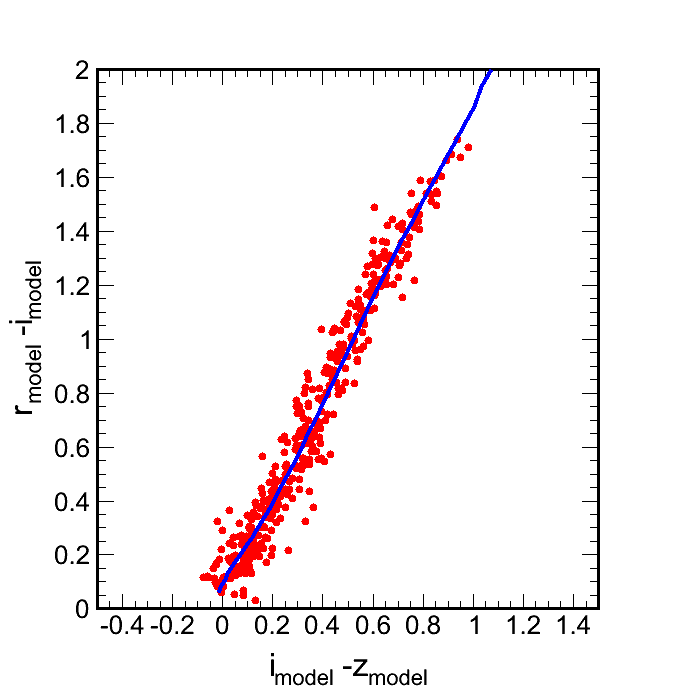}
\vskip-0.0in
\includegraphics[width=0.40\textwidth]{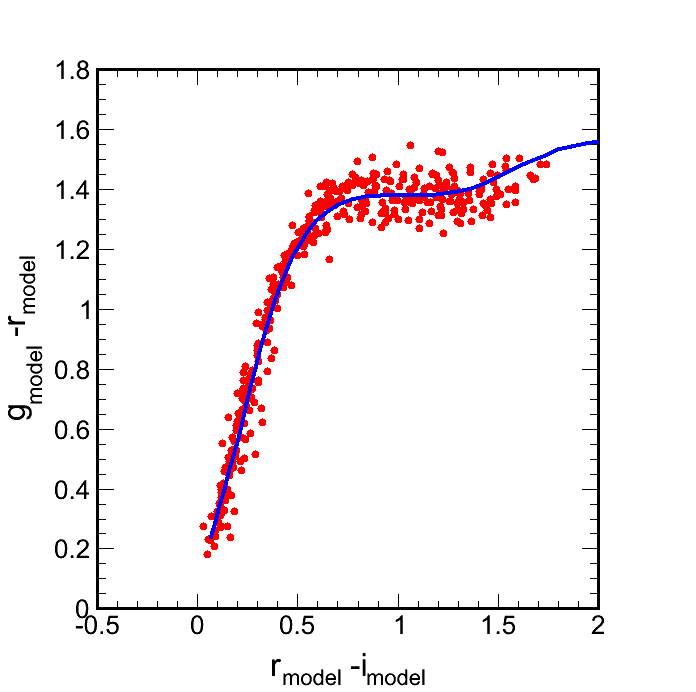}
\vskip-0.0in
\includegraphics[width=0.40\textwidth]{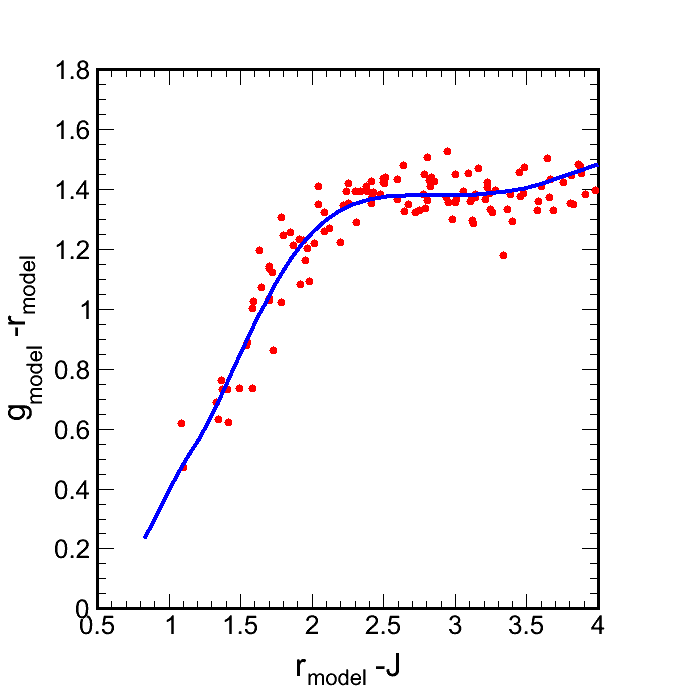}
\caption{The stellar locus in three different color-color spaces for the BCS tile BCS0510-5043.  The blue line shows the expected distribution derived from studies of a large ensemble of stars within the SDSS and 2MASS surveys.  Red points show model magnitudes of stars from the BCS catalogs of this tile. The stellar locus distributions allow us to calibrate the absolute photometry and to assess the quality of the photometry for each tile.}
\label{fig:slrplots-model}
\end{center}
\end{figure}

\subsubsection{Absolute Photometric Calibration}
\label{sec:absolutephotometry}
Once all objects from the coadd are cataloged (in instrumental magnitudes), we proceed to obtain the absolute photometric calibration using Stellar Locus Regression~\citep{high09}. The principle behind this is that the regularity of the stellar main sequence leads to a pre-determined line in color-color space called the stellar locus.  This stellar locus is observed to be invariant over the sky, at least for fields that lie outside the galactic plane.  The constancy of the stellar locus has been used as a cross check of the photometric calibration within the SDSS survey~\citep{ivezic07}.  

Absolute photometric calibration is done after the end of coaddition.  We select star-like objects using a cut on the {\sc SExtractor} {\tt spread\_model} parameter and magnitude error. We then match the observed stars to 2MASS stars from the NOMAD catalog, which is a combination of USNOB and 2MASS datasets, and  which have $JHK$ magnitudes \citep{skrutskie06}.  Color offsets are varied until the observed locus matches the known locus.  Because the 2MASS magnitudes are calibrated with a zeropoint accuracy at the $\sim$2\% level, one can bootstrap the calibration to the other bands.  The known locus is derived using the high quality ``superclean'' SDSS-2MASS matched catalog from~\citet{covey07}.  It consists of $\sim$300,000 high quality stars with data in $ugrizJHK$. A median locus is calculated for each possible color combination in bins of $g-i$.

The fit is done in two stages.  First, a three paramater fit is done to the $g-r$, $r-i$, $i-z$ colors.  Another fit is done using $g-r$ and $r-J$ where the shift in the $g-r$ color is fixed from the first fit.  The first fit provides an accurate calibration of the colors and the second fit fixes the absolute scale.  The fit is done this way because only a fraction of the stars that overlap with 2MASS are saturated in all bands.  We perform the stellar locus calibration for model and 3~arcsec aperture magnitudes, separately.  The model magnitude calibration is then used to calibrate the other magnitudes in the catalog (except for the 3~arcsec magnitudes). The calibration of the 3~arcsec aperture magnitudes determines a PSF dependent aperture correction for the mag\_aper\_3 magnitudes only.  We have found these small aperture magnitudes to provide higher signal to noise colors for faint galaxies in comparison to mag\_model and mag\_auto.

\begin{figure}
\begin{center}
\includegraphics[width=0.49\textwidth]{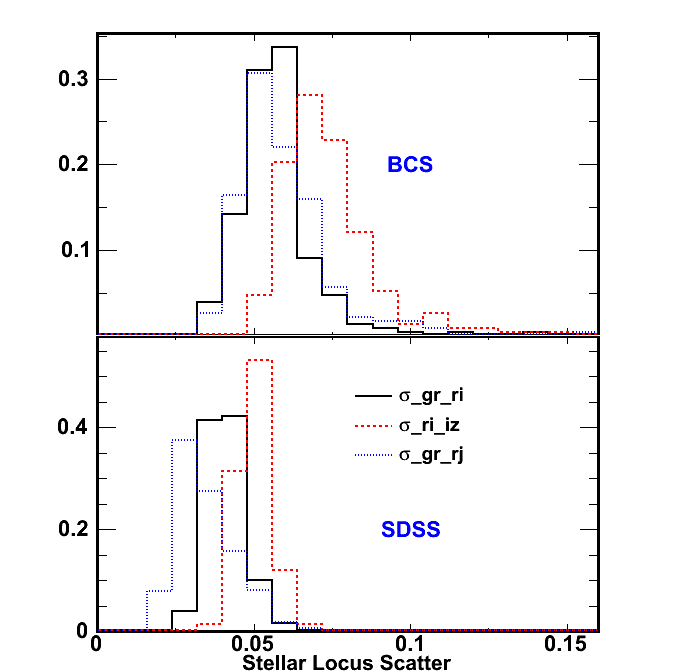}
\caption{Stellar locus scatter (above)  for three color combinations for all tiles in BCS survey (top panel) and the same for SDSS (bottom panel).  Typical BCS scatter is in the 5\% to 8\% range, and offsets after calibration are characteristically 1\% or less. Typical scatter and offsets in the SDSS dataset are smaller than in the BCS survey, reflecting the tighter requirements on photometric quality in SDSS.}
\label{slrfits}
\end{center}
\end{figure}

An example of the stellar locus fits for one coadd tile is shown in Fig.~\ref{fig:slrplots-model}.  Red points are the observed colors of the stars, and the blue line is the median SDSS-2MASS locus.  The orthogonal scatter about the stellar locus for all 3 color combinations for BCS tiles is shown in Fig ~\ref{slrfits}. The rms orthogonal scatter about the stellar locus in  $(g-r, r-i)$, $(r-i, i-z)$ and $(g-r, r-j)$ is 0.059, 0.061 and 0.075 respectively.  Given the scatter and the number of stars available for calibration we can determine the zeropoints in our bands with sub-percent accuracy.

\subsubsection{Testing Stellar Locus Calibration in SDSS}
To validate our photometric calibration algorithms, we applied exactly the same procedure to the full SDSS-2MASS catalog in~\citep{covey07}.  This catalog includes noiser objects than the catalog we used to derive the median stellar locus.  We selected 4 areas (between RA of $120^{\circ}$ and $350^{\circ}$) and divided each into $1^{\circ} \times 1^{\circ}$  patches. We match the objects to obtain 2MASS magnitudes and then apply the same calibration procedure as we did for the BCS catalogs.  The rms scatter distributions for all the three color combinations can be found in Fig.~\ref{slrfits} (bottom panel). The corresponding scatter for SDSS in ($g-r,r-i$), ($r-i,i-z$), and ($g-r,r-j$) is about 0.041, 0.035 and 0.05 respectively and is about 1.5 times smaller than for the BCS catalogs.  This is clear evidence for higher scatter in our stellar photometry as compared to the SDSS photometry.  Assuming this additional source of scatter adds in quadrature with the SDSS observed scatter, we estimate the extra noise in BCS color combinations compared to SDSS  is 0.039, 0.054 and 0.048 in  ($g-r,r-i$), ($r-i,i-z$), and ($g-r,r-j$) respectively.  Because these noise sources are getting contributions from each color, we can estimate that the noise floors are $\delta (g-r)\sim0.027$, $\delta (r-i)\sim0.038$, $\delta (i-z)\sim0.038$.  These then imply noise floors in the stellar photometry within $griz$ bands of approximately 1.9\%, 2.3\%, 2.7\% and 2.7\%, respectively.  This is then in good agreement with the typical repeatability scatter seen in these bands  (see Fig.~\ref{fig:repeatability}) when one considers that $g$ and $r$ bands each have two overlapping exposures and $i$ and $z$ band each have three.

\subsubsection{Star-Galaxy Classification}
\label{subsec:sgseparation}
\begin{figure}
\begin{center}
\includegraphics[width=0.49\textwidth]{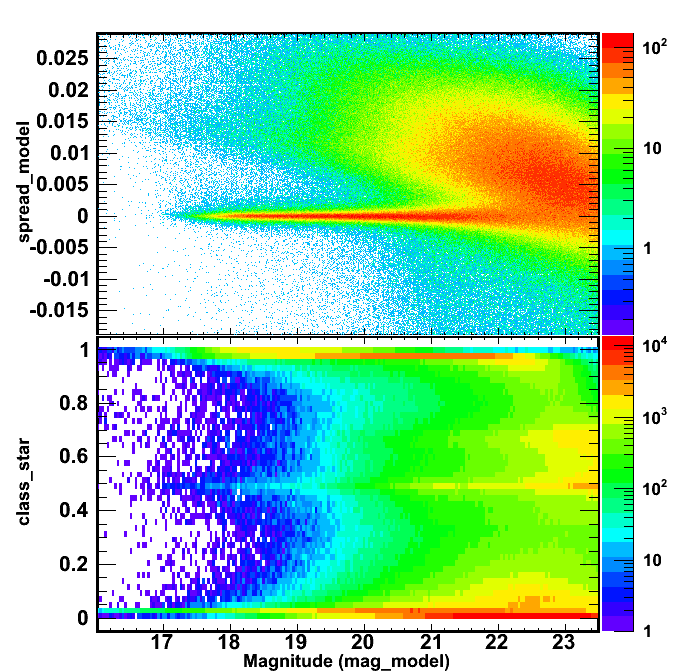}
\caption{Plots of {\tt spread\_model} (top panel) and {\tt class\_star} (bottom panel) as a function of $i$-band magnitudes for the full BCS catalog.  Note that both measurements exhibit separate sequences for stars and galaxies, and that as one moves to fainter magnitudes these sequences merge.  This is simply due to low signal to noise objects not containing enough morphological information for a reliable classification.  However, note also that the new {\tt spread\_model} retains good capability of separating galaxies from stars to fainter magnitudes than {\tt class\_star}.}
\label{sgplots}
\end{center}
\end{figure}

Our current catalogs contain two  star-galaxy classification parameters provided by {\tt SExtractor} :  {\tt class\_star} which has been extensively studied and {\tt spread\_model}, which has been newly developed as part of the DESDM development program. In order  to test their performance and range of magnitudes up to which these measures can be reliably used, we plot the behavior of these two classifiers in $i$-band as a function of $mag\_model$ in Fig.~\ref{sgplots}. {\tt class\_star} lies in the range from 0 to 1.  At bright magnitudes one can see two sequences in {\tt class\_star}  for galaxies and stars near 0 and 1 respectively. The two sequences begin merging as bright as $i=20$ and are significantly merged beyond $i=22$. As described in Sec.~\ref{sec:cataloging}  {\tt spread\_model} uses the local PSF model to quantify the differences between PSF-like objects and resolved objects.  In the {\tt spread\_model} panel it's clear that there is a strong stellar sequence around the value 0.0, and that galaxies exhibit more positive values.  The narrow stellar sequence and the broad galaxy sequence begin merging at $i=22$ in the BCS, but there is significant separation in the two distributions of points down to $i=23$.  Along with  {\tt spread\_model} comes a measurement uncertainty, and so it is, for example, possible to define a sample of objects that lie off the stellar sequence in a statically significant way.  For the BCS data, a good cut to separate stars would be  ${\tt spread\_model}<0.003$.  Detailed studies of this new classification tool have been carried out within the DESDM project and will be carried out elsewhere.

\subsubsection{Quality Control and Science Ready Catalogs}

During the processing within the DESDM system a variety of quality checks are carried out.  These include, for example, thresholding checks on the fraction of flagged pixels within an image and the $\chi^2$ and number of stars used in the astrometric fit of each exposure.  In addition, the system is set up to report on the similarity between correction images (bias, flat, illum, and fringe) against stored templates that have been fully vetted.  During the BCS processing this last facility was not used.

\begin{figure}
\begin{center}
\includegraphics[width=0.49\textwidth]{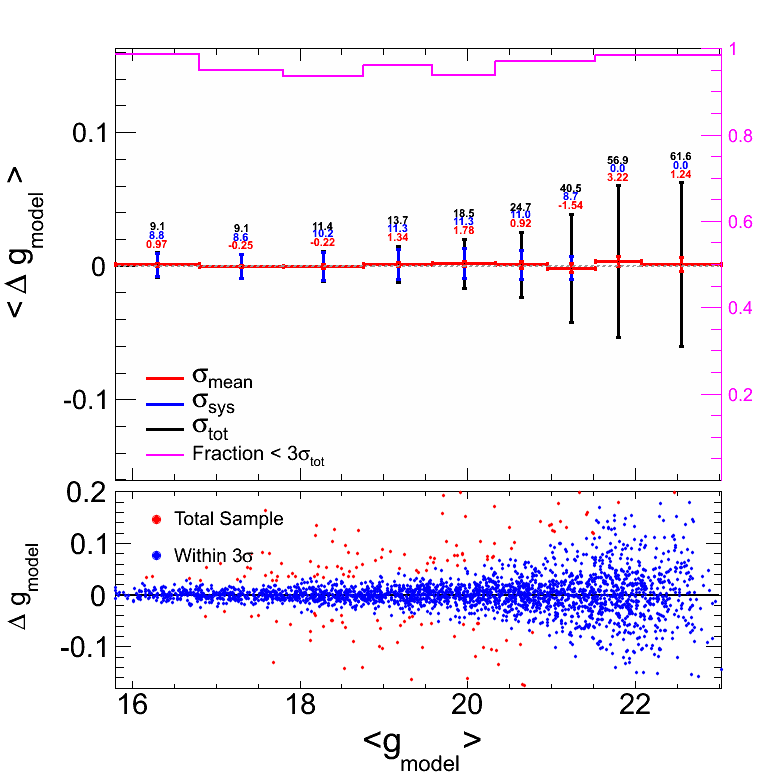}
\caption{Repeatability plots for single-epoch images for BCS tile BCS0549-5043 in $g$-band.  The repeatability is used to test the quality of the photometry in each band and tile. The top panel shows the mean magnitude difference between different single-epoch images which cover same region of sky, binned as a function of magnitude  along with statistical and systematic errors. The bottom panel shows an un-binned representation of the same. Characteristic scatter on the bright end (i.e. the systematic floor) is 2\% to 3\% for $g$ and $r$ and 3\% to 4\% for $i$ and $z$.}
\label{fig:repeatability}
\end{center}
\end{figure}

Our experience has been that problems at any level of processing are most likely to show up in the stages of relative and absolute photometric calibration.  Therefore, for the BCS processing done here we capture a range of photometric quality tests including the number of stars used in the stellar locus calibration and the rms scatter about the true stellar locus of the calibrated data (see Fig.~\ref{fig:slrplots-model}).  In addition, we examine the photometric repeatability for common objects within overlapping images contributing to each tile.  In Fig.~\ref{fig:repeatability} we show an example for the $g$-band in tile BCS0549-5043. This shows the magnitude difference between pairs of overlapping objects versus the average magnitude.  The scatter here includes both statistical and systematic contributions, and the envelope of scatter grows toward faint magnitudes, as expected.  Outlier rejection is done on the point distribution, and all 3$\sigma$ outliers are filtered out and colored red.  In the top panel we plot the mean and rms as well as the outlier fraction of these repeatability distributions as a function of magnitude.  The mean and rms numbers are listed in milli-mags.  Also, the statistical uncertainties of the model magnitudes are used to estimate the systematic magnitude error contribution to the rms.  On the bright end where the statistical noise is very small, the systematic contribution to the rms is close to the total, which is 10~millimags in this case.  As one moves toward the faint end the statistical contribution increases and the estimated systematic contribution plays only a small role in explaining the scatter.  This is just how we expect the photometry to behave. 

Repeatability plots indicate systematic contributions to the photometric errors at the 10 to 20 milli-mag levels for typical $g$ and $r$ band tiles.  For $i$ and $z$ band tiles the systematic noise is closer to 30 to 40 milli-mags.  For all the BCS tiles we have examined these repeatability and stellar locus plots to probe whether the scatter is in acceptable ranges.  In cases where tiles didn't meet these quality control tests we worked on the relative and absolute photometric calibration to improve the data.  In addition to these photometry tests we examined the sky distribution of cataloged objects within each tile.  In cases where large numbers of faint ``junk'' objects were found we attempted to remove them by adjusting the cataloging.  At present all our tiles meet these quality tests except for a handful of tiles that are marked in red in Fig.~\ref{fig:bcscoverage}.  This includes 4 tiles in the 5~hr field and 6 tiles in the 23~hr field, corresponding to $\sim$4\% of the 80~deg$^2$ region.  Ideally, we would reimage these regions to obtain better data.

For every BCS night, the detrending pipeline creates three main types of science image files which 
we denote as {\it raw}, {\it red} and  {\it remap}.   The coadd pipeline produces four coadd images per tile for each of the four bands. Once we have calibrated coadd catalogs for all the processed tiles we run a post-processing program to remove duplicate objects near edges of the tiles.  This is necessary because there is a 2~arcmin overlap between neighboring tiles.  The program selects sources that appear in neighboring tiles that lie within in $0.9$~arcsec radius and for each pair it keeps the object that lies farthest from the edge of its tile.  In this way a single, science ready catalog is prepared for each field.  The 23~hr field catalog contains 1,877,088 objects, and the 5~hr field contains 2,952,282 objects with $i$ model magnitude $< 23.5$.  In the next section we review additional tests of the data quality.

\begin{figure}[htb]
\begin{center}
\includegraphics[width=0.49\textwidth]{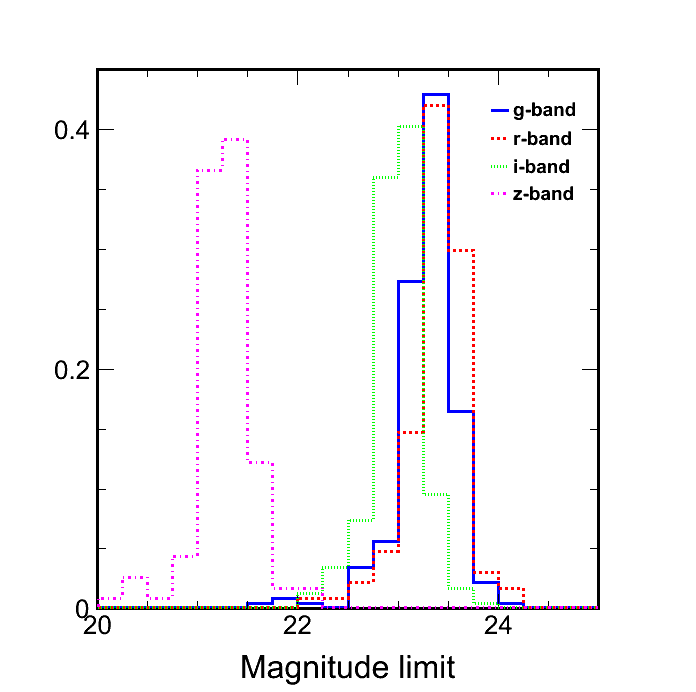}
\caption{Histogram of 10$\sigma$ magnitude limits for all BCS tiles using $mag\_auto$ errors in all four bands.  The median depth values for all BCS tiles are 23.3, 23.4, 23 and 21.3 in $griz$, respectively.  The corresponding 10$\sigma$ point source depths are 23.9, 24.0, 23.6 and 22.1.}
\label{depthwitherrors}
\end{center}
\end{figure}

\subsection{Survey Depth}
\label{sec:bcsdata}

We estimate the 10$\sigma$ photometric depths for galaxies using {\sc SExtractor} $mag\_auto$ errors. This is obtained by doing a linear fit to the relationship between the magnitude and the log of the inverse magnitude error.  As a cross-check we also estimated the depths using information in the weight maps, and the results were comparable.  

The distributions of depth for each band over the full survey is shown in Fig.~\ref{depthwitherrors}.  The median magnitude depths for $griz$ bands are  23.3, 23.4, 23.0 and 21.3 respectively.  These numbers are shallower than the depths we estimated using the NOAO exposure time calculator during the survey planning; those depths were 24.0, 23.9, 23.6, and 22.3.  Our originally proposed depths assume a 2.2~arcsec diameter aperture, whereas galaxies near the 10$\sigma$ detection threshold are typically larger in our images.  We examine the depths of $2''$ aperture photometry and find that the median depths are 24.1, 24.1, 23.5 and 22.2 in $griz$, respectively.  These are within 0.2~mag of our naive estimates, explaining the bulk of the difference.  In addition, we know that during our survey often the conditions were not photometric, and this could introduce another 0.1 to 0.2~mag offset. Another reason for the difference is that the calibrated observed magnitudes also include a correction for galactic extinction and reddening, whereas the estimated depths did not have extinction corrections included.

Corresponding $10\sigma$ point source depths are extracted using model fitting  $mag\_psf$ uncertainties.  The results in bands $griz$ are 23.9, 24.0, 23.6 and 22.1, respectively.  These are in better agreement with the small aperture photometry depths we used to estimate the exposure times for the survey.

\begin{figure}[htb]
\begin{center}
\includegraphics[width=0.49\textwidth]{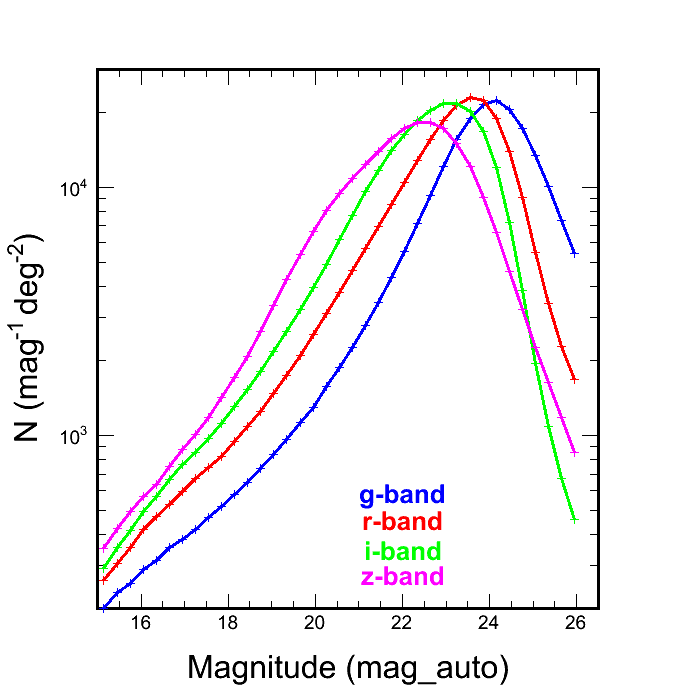}
\caption{Number counts of BCS objects for all four bands in the BCS field using $mag\_auto$.  The turnover magnitudes are 24.15, 23.55, 23.25 and 22.35 in $griz$, respectively.  The corresponding median $mag\_auto$ 10$\sigma$ depths in $griz$ are 23.0, 23.4, 23.0 and 21.3 respectively.}
\label{lognlogs-5hr}
\end{center}
\end{figure}

Another way of probing the depth of the survey is to look at the number counts of sources as a function of magnitude.  Fig.~\ref{lognlogs-5hr} contains the logN-logS from the combined 5~hr and 23~hr fields using $mag\_auto$.  No star-galaxy separation is carried out, because near the detection limit there is not enough morphological information to reliably classify.  The magnitudes of the turnover in the counts corresponds to 24.15, 23.55, 23.25 and 22.35.  These turnover magnitudes mark the onset of significant incompleteness in the catalogs. Estimates of the depth of the 50\% and 90\% completenesss limit for a subset of the tiles appear in~\citet{zenteno11}, but we do not apply that analysis to the whole survey.

\begin{figure}
\begin{center}
\includegraphics[width=0.45\textwidth]{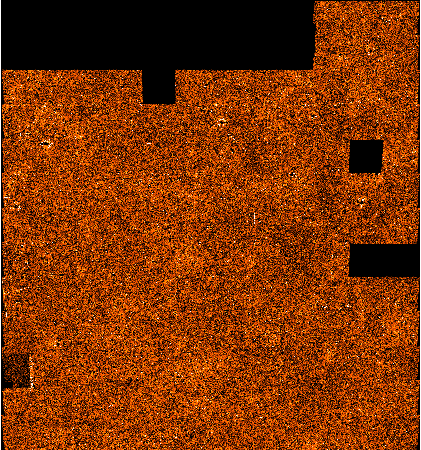}
\vskip0.1in
\includegraphics[width=0.45\textwidth]{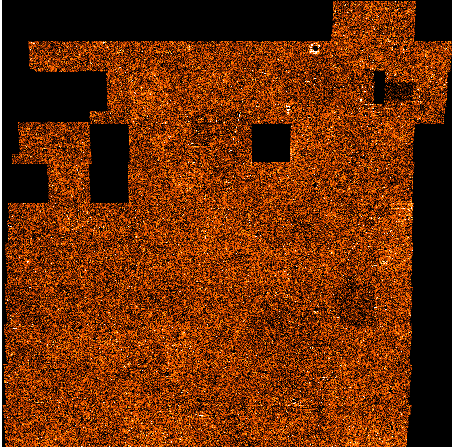}
\caption{Distribution of sources in  5~hr (top)  and 23~hr fields (bottom) from the combined catalogs
after a $mag\_auto$ magnitude cut ($i<22.5$). The gaps show tiles which were not included in the release 
due to data quality problems. Some other tiles have only partial coverage or do not push to the depth of the magnitude cut with good completeness. A logarithmic scale (zscale option in ds9) is used. The uniformity of the source distribution is a demonstration of the photometric uniformity across the survey.}
\label{sourcedistribution}
\end{center}
\end{figure}

Finally, we probe for spatial variations in photometry by examining the distribution of sources above certain flux cuts over the two survey regions.  The distributions of all sources at $i<22.5$ in both the 5~hr and the 23~hr fields are shown in Fig.~\ref{sourcedistribution}.  Objects are excluded for tiles that did not pass our quality tests, and this produces 4 black squares in the 5~hr field, and 6 black squares in the 23~hr field.  The general uniformity of this object density distribution is an indication that the absolute photometric calibration is reasonably consistent across the fields.  In the 5~hr field it is clear that one tile in the lower left does not reach the depth $i=22.5$ reached by the other tiles.  This defect disappears if we examine the density distribution roughly 1 magnitude brighter, indicating this is a depth issue and not a photometric calibration problem.  For the 23~hr field there is a small black rectangular notch in the upper right of the field with an associated dark path.  Within this tile we have verified that too few of our $i$-band exposures met the seeing requirements, and that has led to an uncovered region (the notch) as well as the shadow of lower object density to the right.  Again, this is a depth issue rather than a photometric issue.  There is another shadowed tile visible in the lower right part of the 23~hr field, and this is also a depth issue.

\begin{figure}
\begin{center}
\includegraphics[width=0.45\textwidth]{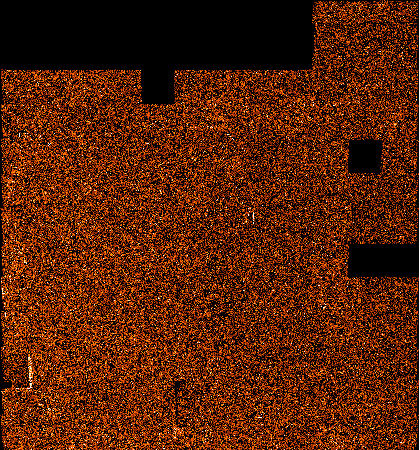}
\vskip0.1in
\includegraphics[width=0.45\textwidth]{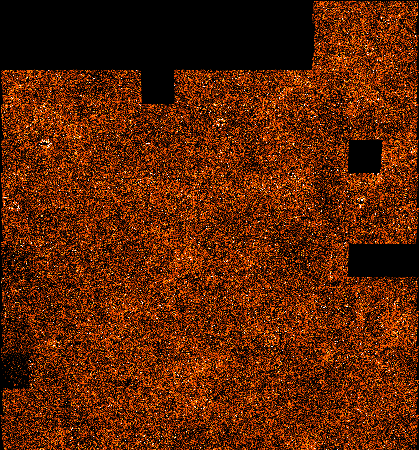}
\caption{Distribution of stars (top) and galaxies (bottom) in  BCS 5~hr field  with $mag\_auto$ $i<22.5$ based on {\tt spread\_model} cut of 0.003. The stars look uniformly distributed and you can see traces of large-scale
structure in galaxy density plot. A logarithmic scale (zscale option in ds9) is used.  We have explored similar plots with {\tt class\_star}, and these contain very large inhomogeneities in the stellar and galaxy distribution, indicating that {\tt spread\_model} offers significant advantages over {\tt class\_star} in the classification of objects in large surveys.}
\label{sgsourcedistribution}
\end{center}
\end{figure}

In Fig.~\ref{sgsourcedistribution} we show similar object density plots for stars and galaxies for the 5~hr field. The stars and galaxies are chosen based on {\tt spread\_model} cut at 0.003, where  all objects with values greater than this threshold were considered galaxies and the rest were considered stars.  A catalog depth cut at $i<22.5$ was imposed. This is shown in Fig.~\ref{sgsourcedistribution}.  The stellar distribution is quite uniform across this field, indicating that {\tt spread\_model} performance is quite robust to variations of PSF across a survey.  Note that the shallow tile in the lower left portion of the survey exhibits edge effects, which we believe are associated with the reduced depth of this tile relative to the others.  In the lower panel is the galaxy distribution.  The same shallow tile shows up in the lower left portion.  In addition, it is clear that the galaxy density is varying as a function of position as expected for the large scale structure of the Universe.  We are quite happy with this performance.  We have explored the same plots in the 23~hr field, and the results are similar.  Moreover, we have explored these plots created using {\tt star\_class} as the classifier.  The spatial distribution is highly inhomogeneous, indicating that {\tt class\_star} cannot be used to reliably separate stars and galaxies in a uniform manner across a large survey.

\subsection{BCS Data release}
We are publicly releasing the BCS catalogs, images and the photo-z training fields
 to the astrophysical community.
All public BCS data products can be downloaded from {\tt http://www.usm.uni-muenchen.de/BCS}.
The BCS catalogs are divided into ascii files for the 5 hour and 23 hour fields.  Separate catalogs are available for the tiles that passed our quality analysis and for the tiles that did not.  Each catalog
contains 63 columns which are described in Table~\ref{tab:bcscatalog}.  We are also making available the coadded images for the BCS survey at the same site.  These images are available in a PSF homogenized form (used for the cataloging) and in the non-homogenized form.  As in the case of the catalogs, we split the tar files by field and by whether the tiles passed our quality tests or did not. These tarballs contain FITS tile compressed images, which reduces the volume by a factor of $\sim$5 relative to the uncompressed coadds.


\section{Photometric Redshifts}

Initial tests of data quality are undertaken by obtaining photometric redshifts
for BCS objects using an artificial neural network.  Neural networks have been
used to determine accurate photometric redshifts in past optical surveys
\citep{Collister07, Oyaizu08a}.  We use \annz, a feed-forward multi-layer
perceptron network designed for finding photometric redshifts
\citep{collister04}.  The network is composed of a series of inputs, several
layers of nodes, and one or more outputs.  Each node is made of a function that
takes its input as a weighted output of each of the previous layer's nodes.
The weights are tuned by training the network on a representative dataset with
known outputs.  The optimal set of weights are those that minimize a cost
function, which reflects the difference between a known output value and the
network's predicted value.

The training process can result in a set of weights that are over-fit to a
particular training set.  Furthermore, a given training process can converge to
a local minimum of the cost function instead of the true minimum.  In \annz,
the first issue is overcome by finding the set of weights that minimizes the
cost function on a separate validation set rather than on the training set
itself.  The second is avoided by training a committee of several networks with
randomized initial weights.  The mean weights from each committee are used in
the final network.  

We train our neural network on 5,820 objects with known redshifts.  It is run
with eight input parameters: four magnitudes $griz$; three colors $g - r$, $r -
i$, and $i - z$; and a concentration index.  $Mag\_auto$ magnitudes are
used for individual filters, $mag\_aper\_3$ magnitudes are used to
determine colors, and the $i$-band {\tt spread\_model} is used for the
concentration index.  Following the guidelines of \citet{Firth03} and
\citet{collister04}, we use a minimally sufficient network architecture and
committee size in hopes of achieving the highest quality results.  We find this
to be a committee of 8 neural networks that each have an architecture of
8:16:16:1 (eight inputs, two hidden layers of 16 nodes each, and one output).
We denote photometric and spectroscopic redshifts as \zphot\ and \zspec,
respectively and have $\Delta z$ represent $\zphot - \zspec$.

\begin{figure}
\begin{center}
\includegraphics[width=0.49\textwidth]{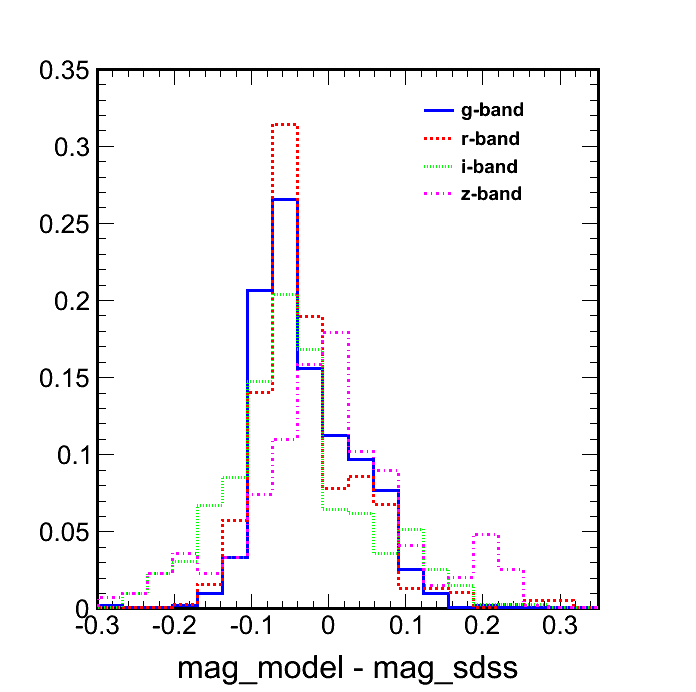}
\caption{Comparison between calibrated model magnitudes for stars from four BCS standard 
tiles (after stacking them  together)  and SDSS magnitudes  after color and extinction corrections for $g$-band. The stars are chosen by requiring that ${\tt class\_star} > 0.8$ in all four bands and also 
{\tt SExtractor} flag $<$ 5. The histograms are normalized to unity. The peak offset between BCS model magnitudes and SDSS is $-0.06$ in $g$, $r$, $i$
and 0.02 in $z$ bands.}
\label{sdsscomp}
\end{center}
\end{figure}

\subsection{Photometry Crosschecks with SDSS}

We compare our photometry with SDSS data by looking at spectroscopic calibration tiles  which overlap with SDSS data and which contain significant numbers of spectroscopic redshifts. As explained in Sec.~\ref{sec:photoz}, these spectroscopic redshifts are then used for training our neural networks to obtained photometric redshifts.  To do a comparison with SDSS catalogs, we applied color and extinction corrections to SDSS catalogs from these tiles. 
The fields which we consider for this purposes are from CNOC and DEEP fields centered at RA, DEC values of (02~hr 25~min, $7^{\circ}$), (02~hr 29~min, $35^{\circ}$), (23~hr 27~min, $8^{\circ}$), and (23~hr 29 min, $12^{\circ}$).

We then do an object by object comparison of colors and magnitudes of all stars from SDSS versus those
 from BCS catalogs in these tiles.  The SDSS magnitudes for objects which overlap BCS tiles go up to 
23.4 in $g$ and  21.6  in $r$, $i$, and $z$.
We consider an object to be matched if it  spatially overlaps to  within $2''$. Since the number of  objects in each tile which overlap with SDSS is small, we combine results from all tiles into one plot for each magnitude or color as necessary. The magnitude comparison for all four bands (using $mag\_model$) is shown in Fig.~\ref{sdsscomp}.  The peak offset between BCS model magnitudes and SDSS magnitudes is approximately $-0.06$ in $g$,$r$ and $i$ bands and about 0.02 in $z$ bands, while the median offset
is $-0.0562$ in $g$, $r$ and $i$ and 0.0087 in $z$.

\begin{figure}
\begin{center}
\includegraphics[width=0.49\textwidth]{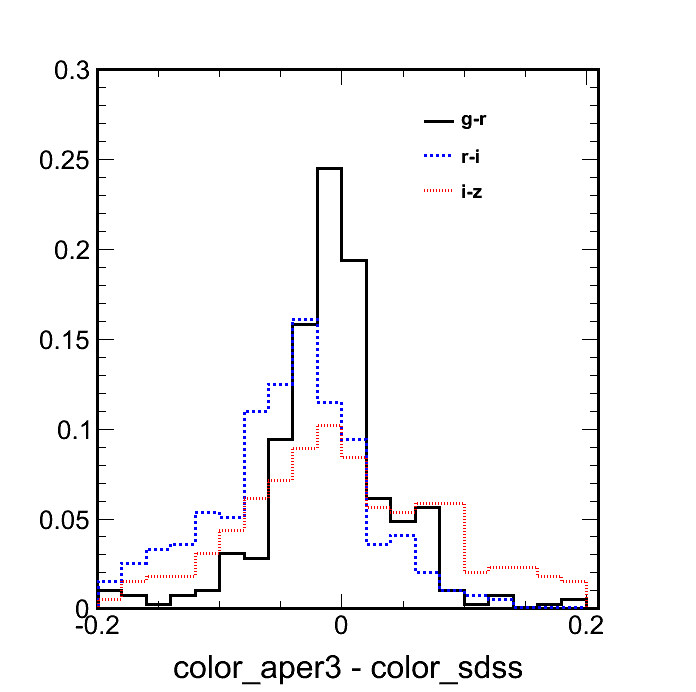}
\caption{Difference in $(g-r)$, $(r-i)$, and $(i-z)$ colors between stars from  BCS tiles and SDSS
using $mag\_aper3$. All cuts are same as in Fig.~\ref{sdsscomp} and histograms are normalized to unity.}
\label{sdsscolor}
\end{center}
\end{figure}

We also do a color comparison using the same cuts for these tiles between BCS and SDSS colors using $mag\_aper3$ (magnitude within a 3 arcsec aperture), because colors are determined using this magnitude in photo-z estimation~(Fig.~\ref{sdsscolor}).  The peak  offset in colors in $g-r$,$r-i$, and $i-z$ is about $-0.01$, $-0.03$, and $-0.02$ magnitudes respectively.  The median offset is about $-0.01$ for $g-r$ and $i-z$ and about $-0.05$ for $r-i$. The rms scatter about the median is 0.052, 0.061 and 0.081 for $g-r$, $r-i$, and $i-z$,  respectively.

\subsection{Photometric Redshift Calibration} \label{sec:photoz}

We obtain our training dataset by dedicating nine of the survey pointings to fields overlapping spectroscopic surveys: CDFS, CFRS, two CNOC2 fields, SSA 22, three DEEP2 fields, and VVDS.  Objects from these fields share their photometric depth and reduction pipeline with the BCS data as well as have known spectroscopic redshifts.  Although this training set is not representative of the survey in sky position, \citet{Abdalla11} show that limiting a neural network training set to small patches of sky does not result in biased redshifts for large surveys.  The key issue is having uniformity of photometry between the training and application fields.

\begin{deluxetable}{lccr}
\tighten
\tabletypesize{\scriptsize}
\tablecaption{Photo-z Training Fields}
\tablewidth{0pt}
\tablehead{\colhead{Survey} & \colhead{RA} & \colhead{Dec} & \colhead{Redshifts}}
ACES\tablenotemark{a} & 03:32 & -27:48 & 2846 \\
CFRS\tablenotemark{b} & 22:17 & 00:91 & 65 \\
CNOC2\tablenotemark{c} & 02:25 & 00:07 & 318 \\
CNOC2\tablenotemark{c} & 02:26 & 00:43 & 164 \\
SSA 22\tablenotemark{d} & 01:40 & 00:01 & 818 \\
DEEP2\tablenotemark{e} & 02:29 & 00:35 & 226 \\
DEEP2\tablenotemark{e} & 23:27 & 00:08 & 414 \\
DEEP2\tablenotemark{e} & 23:29 & 00:12 & 600 \\
VVDS\tablenotemark{f} & 14:00 & 05:00 & 329
\enddata
\tablenotetext{a}{\citet{Cooper11}}
\tablenotetext{b}{\citet{Lilly95a, Lilly95b}}
\tablenotetext{c}{\citet{Yee00}}
\tablenotetext{d}{\citet{Cowie94}}
\tablenotetext{e}{\citet{Davis07, Davis12}}
\tablenotetext{f}{\citet{LeFevre05, LeFevre04}}
\label{tab:training.fields}
\end{deluxetable}

Only objects that have reliable redshifts and photometric parameters are used to train the neural network.  Objects with an $i$-band magnitude $>22.5$ or an $i$-band error $>0.1$ are removed from the training set.  Objects that are unresolved in one or more bands or that have a {\sc SExtractor} flag greater than 2 are removed as well.  Similar cuts are made based on spectroscopic redshift errors, however the nature of the cut varies by catalog.  The DEEP2, CNOC2, and CFRS catalogs provide redshift errors for each measurement.  Objects from these fields are removed if their spectroscopic redshift errors are greater than 0.01.  The ACES catalog (providing coverage of the CDFS field) and the VVDS catalog assign a confidence parameter for each object.  In this case we only include objects with a confidence of 3 or 4 (see respective surveys for definitions).  Both primary and secondary targets from the VVDS survey are included.  Additional cuts were experimented with but found to produce more outliers, a larger sigma, or to reduce the size of the training set too much.

\begin{figure}
\begin{center}
\includegraphics[width=0.49\textwidth]{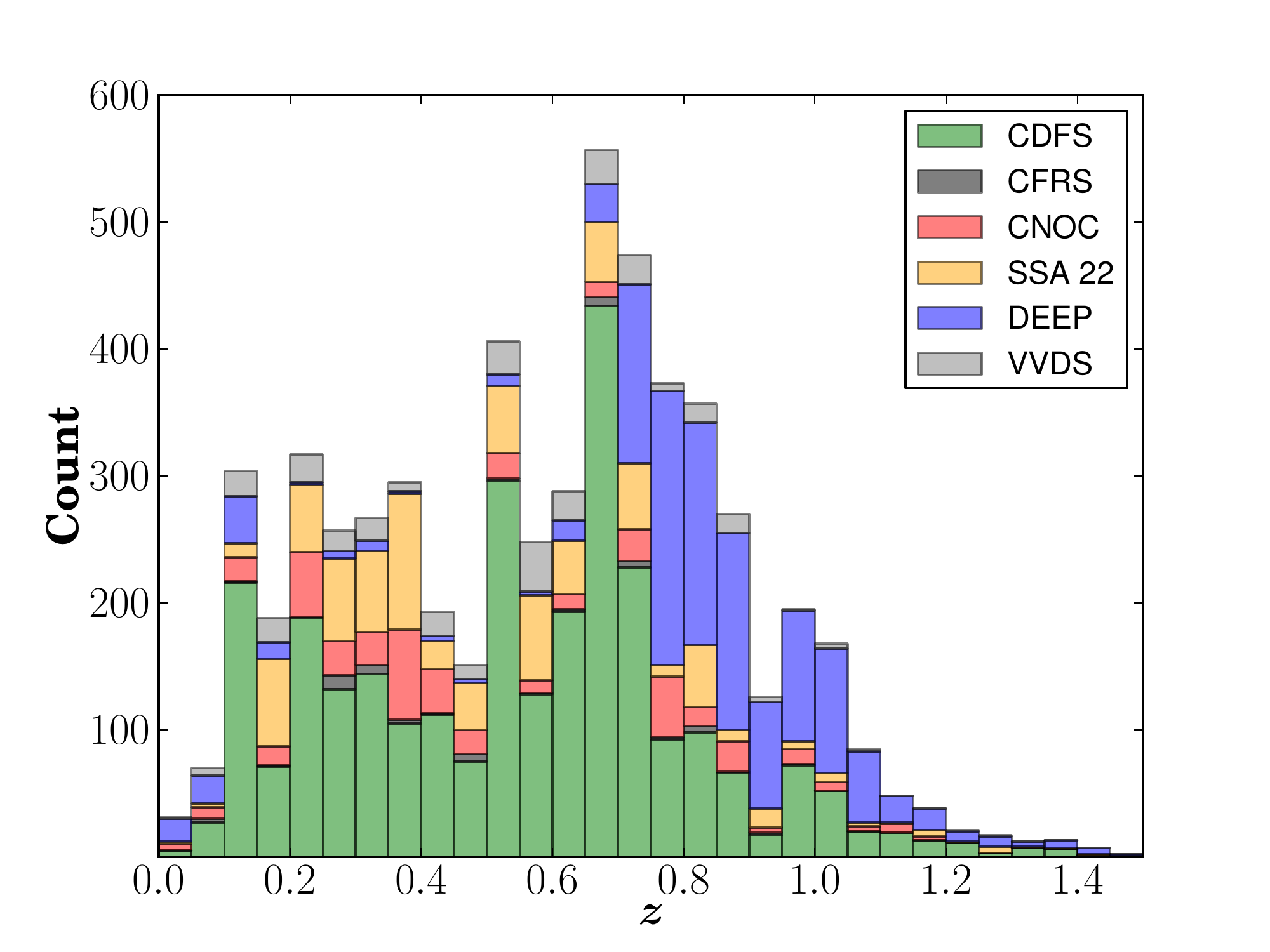}
\caption{Redshift distribution of 5,820 objects from the calibration fields used to train \annz.  The redshift distribution is color coded by source.}
\label{training.hist}
\end{center}
\end{figure}

The final training set contains 5,820 objects.  Table \ref{tab:training.fields} breaks down the number of training objects that pass the filter criteria from each pointing.  Figure \ref{training.hist} further breaks down these objects by redshift bin.  The pointings combine to provide a consistent distribution of redshifts from $0<z\le 1.1$.

We have released the matched catalogs of spectroscopic redshifts along with information from 
BCS catalogs for these fields. This would enable others to develop their own photometric 
redshift estimates using these data.

We evaluate the performance of \annz\ on our data by randomly selecting half of the objects from the training set to train \annz\ while the other half remains for testing.  One sixth of the objects from the training half are removed to form the  validation set (see above).  The result provides 2,910 objects with both photometric and spectroscopic redshifts.

\begin{figure}
\begin{center}
\includegraphics[width=0.49\textwidth]{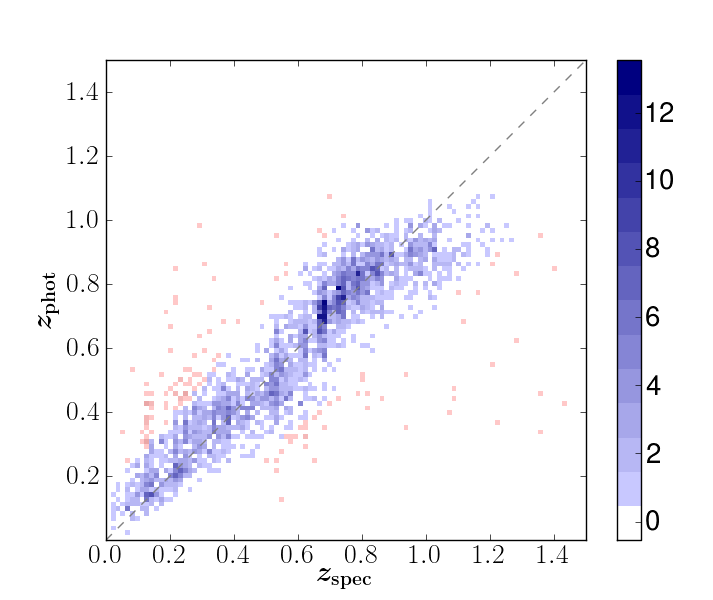}
\includegraphics[width=0.49\textwidth]{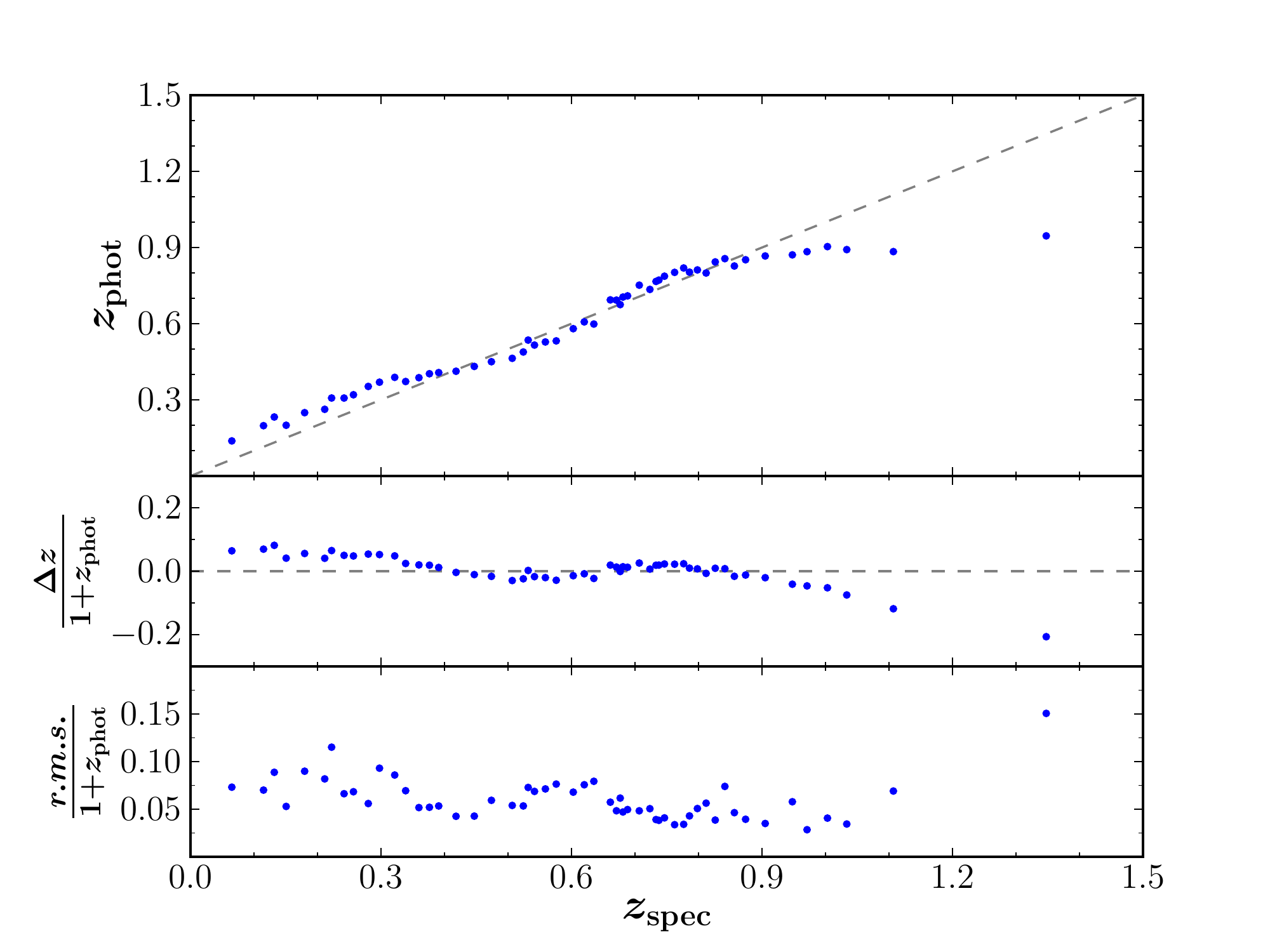}
\caption{ {\it Top panel}: Two-dimensional histogram of \zphot\ vs. \zspec\ for
training set objects that have \zphot\ error $< 0.13$.  Bin sizes are $0.015z
\times 0.015z$.  Red bins count catastrophic outliers as defined by Equation
\ref{catastrophic.outliers}.  Blue bins count all other objects.  5 objects
with \zspec\ or \zphot\ $> 1.5$ are not displayed.  {\it Bottom panel}: The
same training set data are shown with each point representing a bin of 50
objects.}
\label{zspec}
\end{center}
\end{figure}

We measure the photometric redshift performance using three metrics.  The first, following \citet{Ilbert06}, is the normalized median absolute deviation
\begin{equation}
\sigma_{\Delta z/(1+z)} = 1.48 \times {\rm median}\left(\frac{|\Delta z|}{1+z_{\rm spec}}\right). \nonumber
\end{equation}
This metric is better suited for our data than the standard deviation as it is less affected by catastrophic outliers.  The second is the fraction of catastrophic outliers $\eta$ defined as the percentage of objects that satisfy
\begin{equation}
\label{catastrophic.outliers}
\frac{|\Delta z|}{1+z_{\rm spec}} > 0.15.
\end{equation}
The third metric is the net bias in redshift, averaged over all $N$ objects and defined as
\begin{equation}
z_\mathrm{bias} = \frac{1}{N}\sum_{i=1}^N \Delta z_i. \nonumber
\end{equation}

Our training set performs as $\sigma_{\Delta z/(1+z)} = 0.061$ with $\eta =7.49\%$.  Over the entire range of redshifts there is little net bias: $z_\mathrm{bias} = 0.0005$.  These statistics, particularly the fraction of
catastrophic outliers, can be improved by culling objects based on their photometric redshift error.  \annz\ provides redshift errors that are derived from the errors of the input parameters, however there are several other
methods of determining photometric redshift errors.  \citet{Oyaizu08b} evaluate how well various methods improve \zphot\ statistics.  They show that culling objects based on redshift errors derived from magnitude errors are competitive with other methods at reducing the redshift scatter and catastrophic outlier
fraction.

We analyze our photometric performance after culling our data of objects with \zphot\ error $\ge 0.13$ based on errors provided by \annz.  The performance of the culled data improves to $\sigma_{\Delta z/(1+z)} = 0.054$ and $\eta =4.93\%$.  However, the $z_\mathrm{bias}$ increases slightly to 0.0022.  While the bias increases, it is still negligible.  Figure \ref{zspec} demonstrates the performance of \annz\ in determining redshifts.  For objects within the range $0.3 \lesssim \zspec \lesssim0.9$, our photometric and spectroscopic
redshifts match with little bias.  For objects with redshifts below 0.3, there is a positive bias and for objects with redshifts beyond $\zspec \sim 0.9$ there is a negative bias.

\subsection{Application to the full BCS Catalog}

The 5,820 objects from the training set are used to train a committee of 8 \annz\ networks, each with an architecture 8:16:16:1.  This committee is used to determine redshifts and errors for every object in the BCS catalog.  These are included in columns 61 and 62 of the data release.  Because we found a negligible net bias when testing our calibration set, we do not perform a bias correction to redshifts of the BCS catalog.

Many of the objects of the BCS catalog lie outside of the parameter space of data used to train \annz.  While \citet{collister04} have demonstrated success using \annz\ to determine redshifts of galaxies outside the parameter space used to train the network, this was done using a set of galaxies with a very uniform distribution of spectral types.  For the generic distribution of galaxies provided in the BCS catalog, neural networks are unreliable in predicting redshifts outside the trained parameter space. Therefore we indicate whether an object lies inside or outside of the parameter space of the training set with a flag (column 63).  A value of 1 means the object is within the parameter space of the training set and the redshift is reliable.  A value of 0 means the object lays outside the parameter space and the redshift is unreliable.  The flag is based only on the magnitude and magnitude error cuts that were made on the training set (i.e. $i < 22.5$, $i$-error $<0.1$, resolved in all bands).  It is not based on the {\sc SExtractor} flag, the star-galaxy separation criteria, or on photometric redshift errors.

\begin{figure}
\begin{center}
\includegraphics[width=0.49\textwidth]{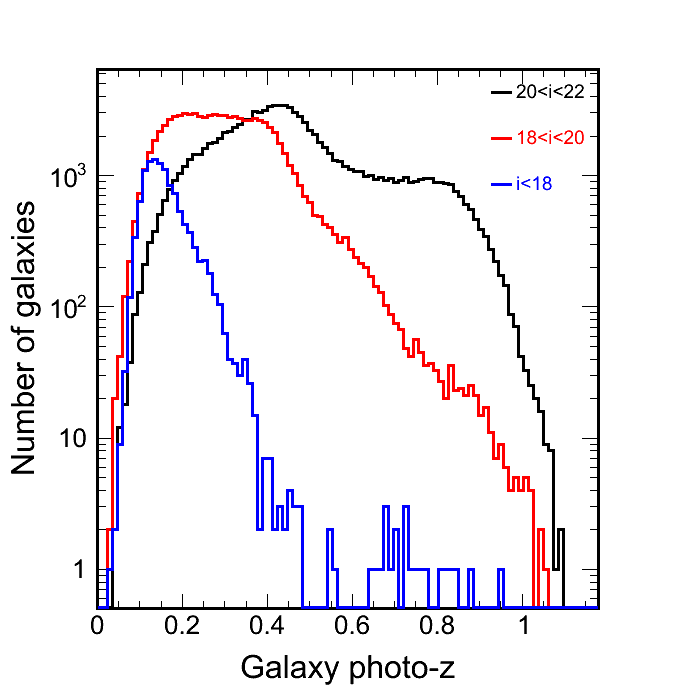}
\caption{Distribution of photometric redshifts of the galaxies that lie within the \annz\ training set parameter space and have \zphot\ error $< 0.05 (1+\zphot)$ and pass the star-galaxy separation test and in different $i$-band model magnitude ranges.
}
\label{BCS.annz.hist}
\end{center}
\end{figure}

We were able to obtain photo-z's for about 1,955,400 objects from the BCS catalog with $i <22$. From these, there
are $\sim$204,600 objects in the catalog that pass the star-galaxy separation criteria in all bands and lie
within the training set parameter space.   The redshift distribution of these BCS objects in different magnitude ranges is shown in Figure \ref{BCS.annz.hist}.  The peak redshift is around $\zphot =0.4$ for $20<i<22$.
 Out of these, there are about 200 objects with $\zphot >1.0$.

Many of the objects that do not pass star-galaxy separation are stars. Since
\annz\ was trained with only galaxies, stellar objects lie outside the
parameter space and therefore do necessarily not get assigned a correct
redshift of $\zphot = 0$. In fact, only a handful of objects in the entire
catalog are assigned a \zphot\ close to zero.  We investigate the performance
of the photo-z's when training a network with both stars and galaxies. Using
the same inputs and network architecture as above but including $\sim$1,000
stars in the training set, \annz\ was successful in assigning stars a redshift
below 0.1 only 70\% of the time. However, redshift assignment of galaxies was
not adversely affected. Only 4 out of approximately 3,400 galaxies were
assigned a redshift less than 0.015. The fraction of catastrophic outliers as
well as $\sigma_{\Delta z/(1+z)}$ were not significantly affected either, so
long as stars are not included in the statistics.  While the results of
training \annz\ with stars are not sufficient to use for the entire BCS
catalog, these preliminary results show some promise.  Furthermore,
\citet{Collister07} have shown better results when training an \annz\ network
specifically for star-galaxy separation. 

\section{Conclusions and Discussion}
\label{sec:conclusion}

In this paper we present  an overview of the Blanco Cosmology Survey (BCS), an $\sim$80~deg$^2$  optical photometric survey in $griz$ bands carried out with the Mosaic2 imager on Blanco 4m telescope  between $2005-2008$.  We discuss the observing strategy within the context of our scientific goals, and we present basic observing characteristics at CTIO such as the sky brightness and delivered image quality.

We provide a detailed description of the data processing, calibration and quality control, which we have carried out using a development version of the Dark Energy Survey Data Management System.  The processing steps in going from raw exposures to science ready catalogs include image detrending and astrometric calibration; this processing is run independently on every night of observations.  This is followed by image co-addition, which combines data from the same region of the sky into deeper coadd images. 

The processing of real data from the Blanco telescope provides a real world stress test of the DESDM system. Many novel algorithmic features, which will be used to process upcoming DES data, were tested on BCS data. These include PSF homogenization, cataloging using PSF corrected, model-fitting photometry, object classification using the new {\tt spread\_model}, absolute photometric calibration using the stellar locus and a variety of quality control tests.  

We present the characteristics of the dataset, including the median estimated $10\sigma$ galaxy photometry depth in the coadds for bands $griz$, which are 23.3, 23.4, 23.0, and 21.3, respectively.  The corresponding point source $10\sigma$ depths in $griz$ are 23.9, 24.0, 23.6, and 22.1, respectively.  We measure the systematic noise floor in our photometry using photometric repeatability in single epoch images and comparisons of the stellar locus scatter from BCS and SDSS.  Both results indicate a noise floor at the $\sim1.9$\% level in $g$,  $\sim2.2$\% in $r$, and $\sim2.7$\% in $i$ and $z$ bands.  This noise floor does not impact the core galaxy cluster science for which the BCS was designed.  We expect that with an improved characterization of the illumination correction using the star flat technique demonstrated in the CFHT Legacy Survey \citep{regnault09} it would be possible to reduce this noise floor further, but given that the current floor is adequate for our science needs we have not included these corrections in our BCS processing.

Our absolute photometric calibration is obtained using the stellar locus and including the 2MASS $J$ band photometry.  We can calibrate our zeropoints at better than $\sim$1\% (statistical) to the stellar locus, and so our overall photometric uniformity is driven by the $\sim$2\% accuracy of the 2MASS survey \citep[e,g,][]{skrutskie06}.  We show that our photometric zeropoint calibration is quite uniform across the survey by showing star and galaxy counts across the survey.  We also demonstrate that with {\tt spread\_model} it is possible to carry out uniform star-galaxy separation even across a large extragalactic survey.

As an additional data quality test, we present photometric redshifts derived from a neural network trained on a sample of objects with spectroscopic redshifts that we targeted during the BCS survey.  The performance of our four band $griz$ photometric redshifts are evaluated based on analysis of a calibration set of over 5,000 galaxies with measured spectroscopic redshifts.  We find good performance with a characteristic scatter of $\sigma_{\Delta z/(1+z)}=0.054$ and an outlier fraction of $\eta=4.93$\%.  Finally, we provided a summary of the output data products from our co-added images and catalogs along with information on how to download them.

Finally, the BCS data have been used for a range of scientific pursuits, which we briefly summarize and reference here to allow the reader to seek additional information as needed. Within the SPT survey, the first four SZE selected clusters were optically confirmed with redshift estimates using BCS data \citep{staniszewski09} and detailed studies of galaxy populations using these clusters were reported in ~\citet{zenteno11}.  The total number of SPT cluster candidates with signal-to-noise ratio $> 4.5$ in BCS footprint is 15~\citep{reichardt12} and among these 10 have been  confirmed with the BCS data and the remaining 5 have redshift lower limits between 1 and 1.5 \citep{song12b}. These clusters and their BCS derived redshifts have figured prominently in SPT publications to date \citep{staniszewski09,vanderlinde10,high10,andersson11,williamson11,reichardt12,song12b}.  The BCS data enabled the serendipitous discovery  of a strong lensing arc of a galaxy at $z = 0.9057$  by a massive galaxy cluster at a redshift of $z=0.3838$~\citep{buckleygeer11}. Additional automated searches for strong lensing arcs have also been carried out, and further analysis of BCS data for weak lensing is in progress.

A sample of about 105 galaxy clusters was found using first three seasons of BCS data using an independent processing~\citep{menanteau09,menanteau10}, and the BCS data were also used for optical confirmation of ACT clusters~\citep{menanteau10b}. Other studies include estimates of weak lensing cluster masses~\citep{mcinnes09} and a search for QSO candidates using $r$-band data~\citep{Jimenez09}.

We used the BCS data to measure photometric redshifts of about 46 X-ray selected clusters in the XMM-BCS survey~\citep{suhada12}.  This X-ray selected sample is currently being used in combination with SPT data to explore the low mass cluster population and its SZE properties (Liu et al., in prep).  In addition, these BCS data are also being used in the analysis of the larger XMM-XXL survey in the 23~hr field (Pierre, private communication).

The BCS data continue to provide an important dataset for SPT.  Recently, the data were used to trace the galaxy populations and were correlated against the SPT CMB-lensing maps \citep{vanengelen12}, demonstrating correlations significant at the 4$\sigma$ to 5$\sigma$ level in both BCS fields~\citep{bleem12}.  The BCS data will provide a valuable optical dataset for combination with a 100~deg$^2$ Spitzer survey over the same region (Stanford, private communication), a 100~deg$^2$ Herschel survey (Carlstrom, private communication), and they will overlap one of the deep mm-wave fields being targeted by SPT-pol (Carlstrom, private communication) until the Dark Energy Survey data are available.

\acknowledgments
We would like to acknowledge Len Cowie for providing us spectroscopic redshifts for objects from the SSA 22 field.  The Munich group acknowledges the support of the Excellence Cluster Universe and from the program TR33:  The Dark Universe, both of which are funded by the Deutsche Forschungs Gemeinschaft.  We acknowledge support from the National Science Foundation (NSF) through grants NSF AST 05-07688, NSF AST 07-08539, NSF AST 07-15036, and NSF AST 08-13534. We acknowledge the support of the University of Illinois where this project was begun.   This paper includes data gathered with the Blanco 4-meter telescope, located at the Cerro Tololo Inter-American Observatory in Chile, which is part of the U.S. National Optical Astronomy Observatory, which is operated by the Association of Universities for Research in Astronomy (AURA), under contract with the NSF.

{\it Facilities:}
\facility{Blanco (MOSAIC)}

\bibliography{spt,paper}

\begin{thebibliography}{84}
\expandafter\ifx\csname natexlab\endcsname\relax\def\natexlab#1{#1}\fi

\bibitem[{{Abdalla} {et~al.}(2011){Abdalla}, {Banerji}, {Lahav}, \&
  {Rashkov}}]{Abdalla11}
{Abdalla}, F.~B., {Banerji}, M., {Lahav}, O., \& {Rashkov}, V. 2011, \mnras,
  1460

\bibitem[{{Abell}(1958)}]{abell58}
{Abell}, G.~O. 1958, \apjs, 3, 211

\bibitem[{{Abell} {et~al.}(1989){Abell}, {Corwin}, \& {Olowin}}]{abell89}
{Abell}, G.~O., {Corwin}, Jr., H.~G., \& {Olowin}, R.~P. 1989, \apjs, 70, 1

\bibitem[{{Adami} {et~al.}(2010){Adami}, {Durret}, {Benoist}, {Coupon},
  {Mazure}, {Meneux}, {Ilbert}, {Blaizot}, {Arnouts}, {Cappi}, {Garilli},
  {Guennou}, {Lebrun}, {Lef{\`e}vre}, {Maurogordato}, {McCracken}, {Mellier},
  {Slezak}, {Tresse}, \& {Ulmer}}]{Adami2009}
{Adami}, C., {et~al.} 2010, \aap, 509, A81+

\bibitem[{{Andersson} {et~al.}(2011){Andersson}, {Benson}, {Ade}, {Aird},
  {Armstrong}, {Bautz}, {Bleem}, {Brodwin}, {Carlstrom}, {Chang}, {Crawford},
  {Crites}, {de Haan}, {Desai}, {Dobbs}, {Dudley}, {Foley}, {Forman},
  {Garmire}, {George}, {Gladders}, {Halverson}, {High}, {Holder}, {Holzapfel},
  {Hrubes}, {Jones}, {Joy}, {Keisler}, {Knox}, {Lee}, {Leitch}, {Lueker},
  {Marrone}, {McMahon}, {Mehl}, {Meyer}, {Mohr}, {Montroy}, {Murray}, {Padin},
  {Plagge}, {Pryke}, {Reichardt}, {Rest}, {Ruel}, {Ruhl}, {Schaffer}, {Shaw},
  {Shirokoff}, {Song}, {Spieler}, {Stalder}, {Staniszewski}, {Stark}, {Stubbs},
  {Vanderlinde}, {Vieira}, {Vikhlinin}, {Williamson}, {Yang}, {Zahn}, \&
  {Zenteno}}]{andersson11}
{Andersson}, K., {et~al.} 2011, \apj, 738, 48

\bibitem[{{Bertin}(2006)}]{bertin06}
{Bertin}, E. 2006, in Astronomical Society of the Pacific Conference Series,
  Vol. 351, Astronomical Data Analysis Software and Systems XV, ed.
  C.~{Gabriel}, C.~{Arviset}, D.~{Ponz}, \& S.~{Enrique}, 112--+

\bibitem[{{Bertin}(2011)}]{bertin11}
{Bertin}, E. 2011, in Astronomical Society of the Pacific Conference Series,
  Vol. 442, Astronomical Data Analysis Software and Systems XX, ed.
  {I.~N.~Evans, A.~Accomazzi, D.~J.~Mink, \& A.~H.~Rots}, 435--+

\bibitem[{{Bertin} {et~al.}(2002){Bertin}, {Mellier}, {Radovich}, {Missonnier},
  {Didelon}, \& {Morin}}]{bertin02}
{Bertin}, E., {Mellier}, Y., {Radovich}, M., {Missonnier}, G., {Didelon}, P.,
  \& {Morin}, B. 2002, in Astronomical Society of the Pacific Conference
  Series, Vol. 281, Astronomical Data Analysis Software and Systems XI, ed.
  {D.~A.~Bohlender, D.~Durand, \& T.~H.~Handley}, 228--+

\bibitem[{Birkinshaw(1999)}]{birkinshaw99}
Birkinshaw, M. 1999, Physics Reports, 310, 97

\bibitem[{{Biviano}(2008)}]{Biviano2008}
{Biviano}, A. 2008, ArXiv e-prints, 0811.3535

\bibitem[{{Bleem} {et~al.}(2012){Bleem}, {van Engelen}, {Holder}, {Aird},
  {Armstrong}, {Ashby}, {Becker}, {Benson}, {Biesiadzinski}, {Brodwin},
  {Busha}, {Carlstrom}, {Chang}, {Cho}, {Crawford}, {Crites}, {de Haan},
  {Desai}, {Dobbs}, {Dor{\'e}}, {Dudley}, {Geach}, {George}, {Gladders},
  {Gonzalez}, {Halverson}, {Harrington}, {High}, {Holden}, {Holzapfel},
  {Hoover}, {Hrubes}, {Joy}, {Keisler}, {Knox}, {Lee}, {Leitch}, {Lueker},
  {Luong-Van}, {Marrone}, {Martinez-Manso}, {McMahon}, {Mehl}, {Meyer}, {Mohr},
  {Montroy}, {Natoli}, {Padin}, {Plagge}, {Pryke}, {Reichardt}, {Rest}, {Ruhl},
  {Saliwanchik}, {Sayre}, {Schaffer}, {Shaw}, {Shirokoff}, {Spieler},
  {Stalder}, {Stanford}, {Staniszewski}, {Stark}, {Stern}, {Story},
  {Vallinotto}, {Vanderlinde}, {Vieira}, {Wechsler}, {Williamson}, \&
  {Zahn}}]{bleem12}
{Bleem}, L.~E., {et~al.} 2012, ArXiv e-prints, 1203.4808

\bibitem[{{Bruzual} \& {Charlot}(2003)}]{bruzual03}
{Bruzual}, G., \& {Charlot}, S. 2003, \mnras, 344, 1000

\bibitem[{{Buckley-Geer} {et~al.}(2011){Buckley-Geer}, {Lin}, {Drabek},
  {Allam}, {Tucker}, {Armstrong}, {Barkhouse}, {Bertin}, {Brodwin}, {Desai},
  {Frieman}, {Hansen}, {High}, {Mohr}, {Lin}, {Ngeow}, {Rest}, {Smith}, {Song},
  \& {Zenteno}}]{buckleygeer11}
{Buckley-Geer}, E.~J., {et~al.} 2011, \apj, 742, 48

\bibitem[{{Butcher} \& {Oemler}(1984)}]{butcher84}
{Butcher}, H., \& {Oemler}, Jr., A. 1984, \apj, 285, 426

\bibitem[{{Carlstrom} {et~al.}(2002){Carlstrom}, {Holder}, \&
  {Reese}}]{carlstrom02}
{Carlstrom}, J.~E., {Holder}, G.~P., \& {Reese}, E.~D. 2002, \araa, 40, 643

\bibitem[{{Cioni} {et~al.}(2011){Cioni}, {Clementini}, {Girardi}, {Guandalini},
  {Gullieuszik}, {Miszalski}, {Moretti}, {Ripepi}, {Rubele}, {Bagheri},
  {Bekki}, {Cross}, {de Blok}, {de Grijs}, {Emerson}, {Evans}, {Gibson},
  {Gonzales-Solares}, {Groenewegen}, {Irwin}, {Ivanov}, {Lewis}, {Marconi},
  {Marquette}, {Mastropietro}, {Moore}, {Napiwotzki}, {Naylor}, {Oliveira},
  {Read}, {Sutorius}, {van Loon}, {Wilkinson}, \& {Wood}}]{VISTA2011}
{Cioni}, M.-R.~L., {et~al.} 2011, \aap, 527, A116

\bibitem[{{Collister} {et~al.}(2007){Collister}, {Lahav}, {Blake}, {Cannon},
  {Croom}, {Drinkwater}, {Edge}, {Eisenstein}, {Loveday}, {Nichol}, {Pimbblet},
  {de Propris}, {Roseboom}, {Ross}, {Schneider}, {Shanks}, \&
  {Wake}}]{Collister07}
{Collister}, A., {et~al.} 2007, \mnras, 375, 68

\bibitem[{{Collister} \& {Lahav}(2004)}]{collister04}
{Collister}, A.~A., \& {Lahav}, O. 2004, \pasp, 116, 345

\bibitem[{{Cooper} {et~al.}(2011){Cooper}, {Yan}, {Dickinson}, {Juneau},
  {Lotz}, {Newman}, {Papovich}, {Salim}, {Walth}, {Weiner}, \&
  {Willmer}}]{Cooper11}
{Cooper}, M.~C., {et~al.} 2011, ArXiv e-prints, 1112.0312

\bibitem[{{Covey} {et~al.}(2007){Covey}, {Ivezi{\'c}}, {Schlegel},
  {Finkbeiner}, {Padmanabhan}, {Lupton}, {Ag{\"u}eros}, {Bochanski}, {Hawley},
  {West}, {Seth}, {Kimball}, {Gogarten}, {Claire}, {Haggard}, {Kaib},
  {Schneider}, \& {Sesar}}]{covey07}
{Covey}, K.~R., {et~al.} 2007, \aj, 134, 2398

\bibitem[{{Cowie} {et~al.}(1994){Cowie}, {Gardner}, {Hu}, {Songaila}, {Hodapp},
  \& {Wainscoat}}]{Cowie94}
{Cowie}, L.~L., {Gardner}, J.~P., {Hu}, E.~M., {Songaila}, A., {Hodapp}, K.-W.,
  \& {Wainscoat}, R.~J. 1994, \apj, 434, 114

\bibitem[{{Cunha} {et~al.}(2010){Cunha}, {Huterer}, \& {Dor{\'e}}}]{cunha10}
{Cunha}, C., {Huterer}, D., \& {Dor{\'e}}, O. 2010, \prd, 82, 023004

\bibitem[{{Dalal} {et~al.}(2008){Dalal}, {Dor{\'e}}, {Huterer}, \&
  {Shirokov}}]{dalal08}
{Dalal}, N., {Dor{\'e}}, O., {Huterer}, D., \& {Shirokov}, A. 2008, \prd, 77,
  123514

\bibitem[{{Davis} {et~al.}(2007){Davis}, {Guhathakurta}, {Konidaris}, {Newman},
  {Ashby}, {Biggs}, {Barmby}, {Bundy}, {Chapman}, {Coil}, {Conselice},
  {Cooper}, {Croton}, {Eisenhardt}, {Ellis}, {Faber}, {Fang}, {Fazio},
  {Georgakakis}, {Gerke}, {Goss}, {Gwyn}, {Harker}, {Hopkins}, {Huang},
  {Ivison}, {Kassin}, {Kirby}, {Koekemoer}, {Koo}, {Laird}, {Le Floc'h}, {Lin},
  {Lotz}, {Marshall}, {Martin}, {Metevier}, {Moustakas}, {Nandra}, {Noeske},
  {Papovich}, {Phillips}, {Rich}, {Rieke}, {Rigopoulou}, {Salim},
  {Schiminovich}, {Simard}, {Smail}, {Small}, {Weiner}, {Willmer}, {Willner},
  {Wilson}, {Wright}, \& {Yan}}]{Davis07}
{Davis}, M., {et~al.} 2007, \apjl, 660, L1

\bibitem[{{de Jong} {et~al.}(2012){de Jong}, {Verdoes Kleijn}, {Kuijken},
  {Valentijn}, {KiDS}, \& {consortiums}}]{dejong12}
{de Jong}, J.~T.~A., {Verdoes Kleijn}, G.~A., {Kuijken}, K.~H., {Valentijn},
  E.~A., {KiDS}, \& {consortiums}, A.-W. 2012, ArXiv e-prints, 1206.1254

\bibitem[{{Dressler}(1980)}]{dressler80}
{Dressler}, A. 1980, \apj, 236, 351

\bibitem[{{Firth} {et~al.}(2003){Firth}, {Lahav}, \& {Somerville}}]{Firth03}
{Firth}, A.~E., {Lahav}, O., \& {Somerville}, R.~S. 2003, \mnras, 339, 1195

\bibitem[{{Fowler}(2004)}]{act2004}
{Fowler}, J.~W. 2004, in Society of Photo-Optical Instrumentation Engineers
  (SPIE) Conference Series, Vol. 5498, Society of Photo-Optical Instrumentation
  Engineers (SPIE) Conference Series, ed. {C.~M.~Bradford, P.~A.~R.~Ade,
  J.~E.~Aguirre, J.~J.~Bock, M.~Dragovan, L.~Duband, L.~Earle, J.~Glenn,
  H.~Matsuhara, B.~J.~Naylor, H.~T.~Nguyen, M.~Yun, \& J.~Zmuidzinas}, 1--10

\bibitem[{{Gal} {et~al.}(2009){Gal}, {Lopes}, {de Carvalho}, {Kohl-Moreira},
  {Capelato}, \& {Djorgovski}}]{Gal09}
{Gal}, R.~R., {Lopes}, P.~A.~A., {de Carvalho}, R.~R., {Kohl-Moreira}, J.~L.,
  {Capelato}, H.~V., \& {Djorgovski}, S.~G. 2009, \aj, 137, 2981

\bibitem[{{Gilbank} {et~al.}(2011){Gilbank}, {Gladders}, {Yee}, \&
  {Hsieh}}]{gilbank11}
{Gilbank}, D.~G., {Gladders}, M.~D., {Yee}, H.~K.~C., \& {Hsieh}, B.~C. 2011,
  \aj, 141, 94

\bibitem[{{Gladders} {et~al.}(2007){Gladders}, {Yee}, {Majumdar}, {Barrientos},
  {Hoekstra}, {Hall}, \& {Infante}}]{gladders07}
{Gladders}, M.~D., {Yee}, H.~K.~C., {Majumdar}, S., {Barrientos}, L.~F.,
  {Hoekstra}, H., {Hall}, P.~B., \& {Infante}, L. 2007, \apj, 655, 128

\bibitem[{{Gonzalez} {et~al.}(2001){Gonzalez}, {Zaritsky}, {Dalcanton}, \&
  {Nelson}}]{apex2010}
{Gonzalez}, A.~H., {Zaritsky}, D., {Dalcanton}, J.~J., \& {Nelson}, A. 2001,
  \apjs, 137, 117

\bibitem[{{Haiman} {et~al.}(2001){Haiman}, {Mohr}, \& {Holder}}]{haiman01}
{Haiman}, Z., {Mohr}, J.~J., \& {Holder}, G.~P. 2001, \apj, 553, 545

\bibitem[{{Hansen} {et~al.}(2009){Hansen}, {Sheldon}, {Wechsler}, \&
  {Koester}}]{hansen09}
{Hansen}, S.~M., {Sheldon}, E.~S., {Wechsler}, R.~H., \& {Koester}, B.~P. 2009,
  \apj, 699, 1333

\bibitem[{{High} {et~al.}(2010){High}, {Stalder}, {Song}, {Ade}, {Aird},
  {Allam}, {Armstrong}, {Barkhouse}, {Benson}, {Bertin}, {Bhattacharya},
  {Bleem}, {Brodwin}, {Buckley-Geer}, {Carlstrom}, {Challis}, {Chang},
  {Crawford}, {Crites}, {de Haan}, {Desai}, {Dobbs}, {Dudley}, {Foley},
  {George}, {Gladders}, {Halverson}, {Hamuy}, {Hansen}, {Holder}, {Holzapfel},
  {Hrubes}, {Joy}, {Keisler}, {Lee}, {Leitch}, {Lin}, {Lin}, {Loehr}, {Lueker},
  {Marrone}, {McMahon}, {Mehl}, {Meyer}, {Mohr}, {Montroy}, {Morell}, {Ngeow},
  {Padin}, {Plagge}, {Pryke}, {Reichardt}, {Rest}, {Ruel}, {Ruhl}, {Schaffer},
  {Shaw}, {Shirokoff}, {Smith}, {Spieler}, {Staniszewski}, {Stark}, {Stubbs},
  {Tucker}, {Vanderlinde}, {Vieira}, {Williamson}, {Wood-Vasey}, {Yang},
  {Zahn}, \& {Zenteno}}]{high10}
{High}, F.~W., {et~al.} 2010, \apj, 723, 1736

\bibitem[{{High} {et~al.}(2009){High}, {Stubbs}, {Rest}, {Stalder}, \&
  {Challis}}]{high09}
{High}, F.~W., {Stubbs}, C.~W., {Rest}, A., {Stalder}, B., \& {Challis}, P.
  2009, \aj, 138, 110

\bibitem[{{Holder} {et~al.}(2001){Holder}, {Haiman}, \& {Mohr}}]{holder01b}
{Holder}, G., {Haiman}, Z., \& {Mohr}, J.~J. 2001, \apjl, 560, L111

\bibitem[{{Ichiki} \& {Takada}(2011)}]{numass}
{Ichiki}, K., \& {Takada}, M. 2011, ArXiv e-prints, 1108.4688

\bibitem[{{Ilbert} {et~al.}(2006){Ilbert}, {Arnouts}, {McCracken},
  {Bolzonella}, {Bertin}, {Le F{\`e}vre}, {Mellier}, {Zamorani}, {Pell{\`o}},
  {Iovino}, {Tresse}, {Le Brun}, {Bottini}, {Garilli}, {Maccagni}, {Picat},
  {Scaramella}, {Scodeggio}, {Vettolani}, {Zanichelli}, {Adami}, {Bardelli},
  {Cappi}, {Charlot}, {Ciliegi}, {Contini}, {Cucciati}, {Foucaud}, {Franzetti},
  {Gavignaud}, {Guzzo}, {Marano}, {Marinoni}, {Mazure}, {Meneux}, {Merighi},
  {Paltani}, {Pollo}, {Pozzetti}, {Radovich}, {Zucca}, {Bondi}, {Bongiorno},
  {Busarello}, {de La Torre}, {Gregorini}, {Lamareille}, {Mathez}, {Merluzzi},
  {Ripepi}, {Rizzo}, \& {Vergani}}]{Ilbert06}
{Ilbert}, O., {et~al.} 2006, \aap, 457, 841

\bibitem[{{Ivezi{\'c}} {et~al.}(2007){Ivezi{\'c}}, {Smith}, {Miknaitis}, {Lin},
  {Tucker}, {Lupton}, {Gunn}, {Knapp}, {Strauss}, {Sesar}, {Doi}, {Tanaka},
  {Fukugita}, {Holtzman}, {Kent}, {Yanny}, {Schlegel}, {Finkbeiner},
  {Padmanabhan}, {Rockosi}, {Juri{\'c}}, {Bond}, {Lee}, {Stoughton}, {Jester},
  {Harris}, {Harding}, {Morrison}, {Brinkmann}, {Schneider}, \&
  {York}}]{ivezic07}
{Ivezi{\'c}}, {\v Z}., {et~al.} 2007, \aj, 134, 973

\bibitem[{{Jimenez} {et~al.}(2009){Jimenez}, {Spergel}, {Niemack}, {Menanteau},
  {Hughes}, {Verde}, \& {Kosowsky}}]{Jimenez09}
{Jimenez}, R., {Spergel}, D.~N., {Niemack}, M.~D., {Menanteau}, F., {Hughes},
  J.~P., {Verde}, L., \& {Kosowsky}, A. 2009, \apjs, 181, 439

\bibitem[{{Koester} {et~al.}(2007){Koester}, {McKay}, {Annis}, {Wechsler},
  {Evrard}, {Bleem}, {Becker}, {Johnston}, {Sheldon}, {Nichol}, {Miller},
  {Scranton}, {Bahcall}, {Barentine}, {Brewington}, {Brinkmann}, {Harvanek},
  {Kleinman}, {Krzesinski}, {Long}, {Nitta}, {Schneider}, {Sneddin}, {Voges},
  \& {York}}]{Koester2007}
{Koester}, B.~P., {et~al.} 2007, \apj, 660, 239

\bibitem[{{Le F{\`e}vre} {et~al.}(2005){Le F{\`e}vre}, {Vettolani}, {Garilli},
  {Tresse}, {Bottini}, {Le Brun}, {Maccagni}, {Picat}, {Scaramella},
  {Scodeggio}, {Zanichelli}, {Adami}, {Arnaboldi}, {Arnouts}, {Bardelli},
  {Bolzonella}, {Cappi}, {Charlot}, {Ciliegi}, {Contini}, {Foucaud},
  {Franzetti}, {Gavignaud}, {Guzzo}, {Ilbert}, {Iovino}, {McCracken}, {Marano},
  {Marinoni}, {Mathez}, {Mazure}, {Meneux}, {Merighi}, {Paltani}, {Pell{\`o}},
  {Pollo}, {Pozzetti}, {Radovich}, {Zamorani}, {Zucca}, {Bondi}, {Bongiorno},
  {Busarello}, {Lamareille}, {Mellier}, {Merluzzi}, {Ripepi}, \&
  {Rizzo}}]{LeFevre05}
{Le F{\`e}vre}, O., {et~al.} 2005, \aap, 439, 845

\bibitem[{{Le F{\`e}vre} {et~al.}(2004){Le F{\`e}vre}, {Vettolani}, {Paltani},
  {Tresse}, {Zamorani}, {Le Brun}, {Moreau}, {Bottini}, {Maccagni}, {Picat},
  {Scaramella}, {Scodeggio}, {Zanichelli}, {Adami}, {Arnouts}, {Bardelli},
  {Bolzonella}, {Cappi}, {Charlot}, {Contini}, {Foucaud}, {Franzetti},
  {Garilli}, {Gavignaud}, {Guzzo}, {Ilbert}, {Iovino}, {McCracken}, {Mancini},
  {Marano}, {Marinoni}, {Mathez}, {Mazure}, {Meneux}, {Merighi}, {Pell{\`o}},
  {Pollo}, {Pozzetti}, {Radovich}, {Zucca}, {Arnaboldi}, {Bondi}, {Bongiorno},
  {Busarello}, {Ciliegi}, {Gregorini}, {Mellier}, {Merluzzi}, {Ripepi}, \&
  {Rizzo}}]{LeFevre04}
------. 2004, \aap, 428, 1043

\bibitem[{{Lilly} {et~al.}(1995{\natexlab{a}}){Lilly}, {Hammer}, {Le Fevre}, \&
  {Crampton}}]{Lilly95b}
{Lilly}, S.~J., {Hammer}, F., {Le Fevre}, O., \& {Crampton}, D.
  1995{\natexlab{a}}, \apj, 455, 75

\bibitem[{{Lilly} {et~al.}(1995{\natexlab{b}}){Lilly}, {Le Fevre}, {Crampton},
  {Hammer}, \& {Tresse}}]{Lilly95a}
{Lilly}, S.~J., {Le Fevre}, O., {Crampton}, D., {Hammer}, F., \& {Tresse}, L.
  1995{\natexlab{b}}, \apj, 455, 50

\bibitem[{{Lin} {et~al.}(2006){Lin}, {Mohr}, {Gonzalez}, \& {Stanford}}]{lin06}
{Lin}, Y., {Mohr}, J.~J., {Gonzalez}, A.~H., \& {Stanford}, S.~A. 2006, \apjl,
  650, L99

\bibitem[{{Lin} {et~al.}(2003){Lin}, {Mohr}, \& {Stanford}}]{lin03b}
{Lin}, Y., {Mohr}, J.~J., \& {Stanford}, S.~A. 2003, \apj, 591, 749

\bibitem[{{Lin} \& {Mohr}(2004)}]{lin04b}
{Lin}, Y.-T., \& {Mohr}, J.~J. 2004, \apj, 617, 879

\bibitem[{{Majumdar} \& {Mohr}(2003)}]{majumdar03}
{Majumdar}, S., \& {Mohr}, J.~J. 2003, \apj, 585, 603

\bibitem[{{Majumdar} \& {Mohr}(2004)}]{majumdar04}
------. 2004, \apj, 613, 41

\bibitem[{{McInnes} {et~al.}(2009){McInnes}, {Menanteau}, {Heavens}, {Hughes},
  {Jimenez}, {Massey}, {Simon}, \& {Taylor}}]{mcinnes09}
{McInnes}, R.~N., {Menanteau}, F., {Heavens}, A.~F., {Hughes}, J.~P.,
  {Jimenez}, R., {Massey}, R., {Simon}, P., \& {Taylor}, A. 2009, \mnras, 399,
  L84

\bibitem[{{Menanteau} {et~al.}(2010{\natexlab{a}}){Menanteau}, {Gonz{\'a}lez},
  {Juin}, {Marriage}, {Reese}, {Acquaviva}, {Aguirre}, {Appel}, {Baker},
  {Barrientos}, {Battistelli}, {Bond}, {Das}, {Deshpande}, {Devlin}, {Dicker},
  {Dunkley}, {D{\"u}nner}, {Essinger-Hileman}, {Fowler}, {Hajian}, {Halpern},
  {Hasselfield}, {Hern{\'a}ndez-Monteagudo}, {Hilton}, {Hincks}, {Hlozek},
  {Huffenberger}, {Hughes}, {Infante}, {Irwin}, {Klein}, {Kosowsky}, {Lin},
  {Marsden}, {Moodley}, {Niemack}, {Nolta}, {Page}, {Parker}, {Partridge},
  {Sehgal}, {Sievers}, {Spergel}, {Staggs}, {Swetz}, {Switzer}, {Thornton},
  {Trac}, {Warne}, \& {Wollack}}]{menanteau10b}
{Menanteau}, F., {et~al.} 2010{\natexlab{a}}, \apj, 723, 1523

\bibitem[{{Menanteau} \& {Hughes}(2009)}]{menanteau09a}
{Menanteau}, F., \& {Hughes}, J.~P. 2009, \apjl, 694, L136

\bibitem[{{Menanteau} {et~al.}(2010{\natexlab{b}}){Menanteau}, {Hughes},
  {Barrientos}, {Deshpande}, {Hilton}, {Infante}, {Jimenez}, {Kosowsky},
  {Moodley}, {Spergel}, \& {Verde}}]{menanteau10}
{Menanteau}, F., {et~al.} 2010{\natexlab{b}}, \apjs, 191, 340

\bibitem[{{Menanteau} {et~al.}(2009){Menanteau}, {Hughes}, {Jimenez},
  {Hernandez-Monteagudo}, {Verde}, {Kosowsky}, {Moodley}, {Infante}, \&
  {Roche}}]{menanteau09}
------. 2009, \apj, 698, 1221

\bibitem[{{Mohr} {et~al.}(2008){Mohr}, {Adams}, {Barkhouse}, {Beldica},
  {Bertin}, {Cai}, {da Costa}, {Darnell}, {Daues}, {Jarvis}, {Gower}, {Lin},
  {Martelli}, {Neilsen}, {Ngeow}, {Ogando}, {Parga}, {Sheldon}, {Tucker},
  {Kuropatkin}, \& {Stoughton}}]{mohr08}
{Mohr}, J.~J., {et~al.} 2008, in Society of Photo-Optical Instrumentation
  Engineers (SPIE) Conference Series, Vol. 7016, Society of Photo-Optical
  Instrumentation Engineers (SPIE) Conference Series

\bibitem[{{Monet} {et~al.}(2003){Monet}, {Levine}, {Canzian}, {Ables}, {Bird},
  {Dahn}, {Guetter}, {Harris}, {Henden}, {Leggett}, {Levison}, {Luginbuhl},
  {Martini}, {Monet}, {Munn}, {Pier}, {Rhodes}, {Riepe}, {Sell}, {Stone},
  {Vrba}, {Walker}, {Westerhout}, {Brucato}, {Reid}, {Schoening}, {Hartley},
  {Read}, \& {Tritton}}]{monet03}
{Monet}, D.~G., {et~al.} 2003, \aj, 125, 984

\bibitem[{{Newman} {et~al.}(2012){Newman}, {Cooper}, {Davis}, {Faber}, {Coil},
  {Guhathakurta}, {Koo}, {Phillips}, {Conroy}, {Dutton}, {Finkbeiner}, {Gerke},
  {Rosario}, {Weiner}, {Willmer}, {Yan}, {Harker}, {Kassin}, {Konidaris},
  {Lai}, {Madgwick}, {Noeske}, {Wirth}, {Connolly}, {Kaiser}, {Kirby},
  {Lemaux}, {Lin}, {Lotz}, {Luppino}, {Marinoni}, {Matthews}, {Metevier}, \&
  {Schiavon}}]{Davis12}
{Newman}, J.~A., {et~al.} 2012, ArXiv e-prints, 1203.3192

\bibitem[{{Ngeow} {et~al.}(2006){Ngeow}, {Mohr}, {Alam}, {Barkhouse},
  {Beldica}, {Cai}, {Daues}, {Plante}, {Annis}, {Lin}, {Tucker}, \&
  {Smith}}]{ngeow06}
{Ngeow}, C., {et~al.} 2006, in Society of Photo-Optical Instrumentation
  Engineers (SPIE) Conference Series, Vol. 6270, Society of Photo-Optical
  Instrumentation Engineers (SPIE) Conference Series

\bibitem[{{Oyaizu} {et~al.}(2008{\natexlab{a}}){Oyaizu}, {Lima}, {Cunha},
  {Lin}, \& {Frieman}}]{Oyaizu08b}
{Oyaizu}, H., {Lima}, M., {Cunha}, C.~E., {Lin}, H., \& {Frieman}, J.
  2008{\natexlab{a}}, \apj, 689, 709

\bibitem[{{Oyaizu} {et~al.}(2008{\natexlab{b}}){Oyaizu}, {Lima}, {Cunha},
  {Lin}, {Frieman}, \& {Sheldon}}]{Oyaizu08a}
{Oyaizu}, H., {Lima}, M., {Cunha}, C.~E., {Lin}, H., {Frieman}, J., \&
  {Sheldon}, E.~S. 2008{\natexlab{b}}, \apj, 674, 768

\bibitem[{{Perlmutter} {et~al.}(1999){Perlmutter}, {Aldering}, {Goldhaber},
  {Knop}, {Nugent}, {Castro}, {Deustua}, {Fabbro}, {Goobar}, {Groom}, {Hook},
  {Kim}, {Kim}, {Lee}, {Nunes}, {Pain}, {Pennypacker}, {Quimby}, {Lidman},
  {Ellis}, {Irwin}, {McMahon}, {Ruiz-Lapuente}, {Walton}, {Schaefer}, {Boyle},
  {Filippenko}, {Matheson}, {Fruchter}, {Panagia}, {Newberg}, {Couch}, \& {The
  Supernova Cosmology Project}}]{perlmutter99a}
{Perlmutter}, S., {et~al.} 1999, \apj, 517, 565

\bibitem[{{Regnault} {et~al.}(2009){Regnault}, {Conley}, {Guy}, {Sullivan},
  {Cuillandre}, {Astier}, {Balland}, {Basa}, {Carlberg}, {Fouchez}, {Hardin},
  {Hook}, {Howell}, {Pain}, {Perrett}, \& {Pritchet}}]{regnault09}
{Regnault}, N., {et~al.} 2009, \aap, 506, 999

\bibitem[{{Reichardt} {et~al.}(2009){Reichardt}, {Ade}, {Bock}, {Bond},
  {Brevik}, {Contaldi}, {Daub}, {Dempsey}, {Goldstein}, {Holzapfel}, {Kuo},
  {Lange}, {Lueker}, {Newcomb}, {Peterson}, {Ruhl}, {Runyan}, \&
  {Staniszewski}}]{acbar}
{Reichardt}, C.~L., {et~al.} 2009, \apj, 694, 1200

\bibitem[{{Reichardt} {et~al.}(2012){Reichardt}, {Stalder}, {Bleem}, {Montroy},
  {Aird}, {Andersson}, {Armstrong}, {Ashby}, {Bautz}, {Bayliss}, {Bazin},
  {Benson}, {Brodwin}, {Carlstrom}, {Chang}, {Cho}, {Clocchiatti}, {Crawford},
  {Crites}, {de Haan}, {Desai}, {Dobbs}, {Dudley}, {Foley}, {Forman}, {George},
  {Gladders}, {Gonzalez}, {Halverson}, {Harrington}, {High}, {Holder},
  {Holzapfel}, {Hoover}, {Hrubes}, {Jones}, {Joy}, {Keisler}, {Knox}, {Lee},
  {Leitch}, {Liu}, {Lueker}, {Luong-Van}, {Mantz}, {Marrone}, {McDonald},
  {McMahon}, {Mehl}, {Meyer}, {Mocanu}, {Mohr}, {Murray}, {Natoli}, {Padin},
  {Plagge}, {Pryke}, {Rest}, {Ruel}, {Ruhl}, {Saliwanchik}, {Saro}, {Sayre},
  {Schaffer}, {Shaw}, {Shirokoff}, {Song}, {Spieler}, {Staniszewski}, {Stark},
  {Story}, {Stubbs}, {Suhada}, {van Engelen}, {Vanderlinde}, {Vieira},
  {Vikhlinin}, {Williamson}, {Zahn}, \& {Zenteno}}]{reichardt12}
------. 2012, ArXiv e-prints, 1203.5775

\bibitem[{{Rest} {et~al.}(2005){Rest}, {Stubbs}, {Becker}, {Miknaitis},
  {Miceli}, {Covarrubias}, {Hawley}, {Smith}, {Suntzeff}, {Olsen}, {Prieto},
  {Hiriart}, {Welch}, {Cook}, {Nikolaev}, {Huber}, {Prochtor}, {Clocchiatti},
  {Minniti}, {Garg}, {Challis}, {Keller}, \& {Schmidt}}]{rest05a}
{Rest}, A., {et~al.} 2005, \apj, 634, 1103

\bibitem[{{Rozo} {et~al.}(2010){Rozo}, {Wechsler}, {Rykoff}, {Annis}, {Becker},
  {Evrard}, {Frieman}, {Hansen}, {Hao}, {Johnston}, {Koester}, {McKay},
  {Sheldon}, \& {Weinberg}}]{rozo10}
{Rozo}, E., {et~al.} 2010, \apj, 708, 645

\bibitem[{{Ruhl} {et~al.}(2004){Ruhl}, {Ade}, {Carlstrom}, {Cho}, {Crawford},
  {Dobbs}, {Greer}, {Halverson}, {Holzapfel}, {Lanting}, {Lee}, {Lerlinitch},
  {Leong}, {Lu}, {Lueker}, {Mehl}, {Meyer}, {Mohr}, {Padin}, {Plagge}, {Pryke},
  {Runyan}, {Schwan}, {Sharp}, {Spieler}, {Staniszewski}, \& {Stark}}]{ruhl04}
{Ruhl}, J., {et~al.} 2004, in Proc. SPIE, Vol. 5498, Millimeter and
  Submillimeter Detectors for Astronomy II, ed. J.~{Zmuidzinas}, W.~S.
  {Holland}, \& S.~{Withington} (Bellingham: SPIE Optical Engineering Press),
  11--29

\bibitem[{{Schmidt} {et~al.}(1998){Schmidt}, {Suntzeff}, {Phillips},
  {Schommer}, {Clocchiatti}, {Kirshner}, {Garnavich}, {Challis}, {Leibundgut},
  {Spyromilio}, {Riess}, {Filippenko}, {Hamuy}, {Smith}, {Hogan}, {Stubbs},
  {Diercks}, {Reiss}, {Gilliland}, {Tonry}, {Maza}, {Dressler}, {Walsh}, \&
  {Ciardullo}}]{schmidt98}
{Schmidt}, B.~P., {et~al.} 1998, \apj, 507, 46

\bibitem[{{Skrutskie} {et~al.}(2006){Skrutskie}, {Cutri}, {Stiening},
  {Weinberg}, {Schneider}, {Carpenter}, {Beichman}, {Capps}, {Chester},
  {Elias}, {Huchra}, {Liebert}, {Lonsdale}, {Monet}, {Price}, {Seitzer},
  {Jarrett}, {Kirkpatrick}, {Gizis}, {Howard}, {Evans}, {Fowler}, {Fullmer},
  {Hurt}, {Light}, {Kopan}, {Marsh}, {McCallon}, {Tam}, {Van Dyk}, \&
  {Wheelock}}]{skrutskie06}
{Skrutskie}, M.~F., {et~al.} 2006, \aj, 131, 1163

\bibitem[{{Smith} {et~al.}(2002){Smith}, {Tucker}, {Kent}, {Richmond},
  {Fukugita}, {Ichikawa}, {Ichikawa}, {Jorgensen}, {Uomoto}, {Gunn}, {Hamabe},
  {Watanabe}, {Tolea}, {Henden}, {Annis}, {Pier}, {McKay}, {Brinkmann}, {Chen},
  {Holtzman}, {Shimasaku}, \& {York}}]{Smith2002}
{Smith}, J.~A., {et~al.} 2002, \aj, 123, 2121

\bibitem[{{Soares-Santos} {et~al.}(2011){Soares-Santos}, {Annis}, {Bonati},
  {Buckley-Geer}, {Cease}, {DePoy}, {Derylo}, {Diehl}, {Elliott}, {Estrada},
  {Finley}, {Flaugher}, {Frieman}, {Hao}, {Honscheid}, {Karliner}, {Krempetz},
  {Kuehn}, {Kuhlmann}, {Kuk}, {Lin}, {Merrit}, {Neilsen}, {Scott}, {Serrano},
  {Shaw}, {Schultz}, {Stuermer}, {Sypniewski}, {Thaler}, {Walker}, {Walton},
  {Wester}, {Yanny}, \& {for the DES Collaboration}}]{decam}
{Soares-Santos}, M., {et~al.} 2011, ArXiv e-prints, 1111.4717

\bibitem[{{Song} {et~al.}(2012){Song}, {A}, \& {B}}]{song12b}
{Song}, J., {A}, \& {B}. 2012, In prep

\bibitem[{{Staniszewski} {et~al.}(2009){Staniszewski}, {Ade}, {Aird}, {Benson},
  {Bleem}, {Carlstrom}, {Chang}, {Cho}, {Crawford}, {Crites}, {de Haan},
  {Dobbs}, {Halverson}, {Holder}, {Holzapfel}, {Hrubes}, {Joy}, {Keisler},
  {Lanting}, {Lee}, {Leitch}, {Loehr}, {Lueker}, {McMahon}, {Mehl}, {Meyer},
  {Mohr}, {Montroy}, {Ngeow}, {Padin}, {Plagge}, {Pryke}, {Reichardt}, {Ruhl},
  {Schaffer}, {Shaw}, {Shirokoff}, {Spieler}, {Stalder}, {Stark},
  {Vanderlinde}, {Vieira}, {Zahn}, \& {Zenteno}}]{staniszewski09}
{Staniszewski}, Z., {et~al.} 2009, \apj, 701, 32

\bibitem[{{Sunyaev} \& {Zel'dovich}(1972)}]{sunyaev72}
{Sunyaev}, R.~A., \& {Zel'dovich}, Y.~B. 1972, Comments on Astrophysics and
  Space Physics, 4, 173

\bibitem[{{Tucker} {et~al.}(2007){Tucker}, {Annis}, {Lin}, {Kent}, {Stoughton},
  {Peoples}, {Allam}, {Mohr}, {Barkhouse}, {Ngeow}, {Alam}, {Beldica}, {Cai},
  {Daues}, {Plante}, {Miller}, {Smith}, \& {Suntzeff}}]{Tucker2007}
{Tucker}, D.~L., {et~al.} 2007, in Astronomical Society of the Pacific
  Conference Series, Vol. 364, The Future of Photometric, Spectrophotometric
  and Polarimetric Standardization, ed. {C.~Sterken}, 187--+

\bibitem[{{{\v S}uhada} {et~al.}(2012){{\v S}uhada}, {Song}, {B{\"o}hringer},
  {Mohr}, {Chon}, {Finoguenov}, {Fassbender}, {Desai}, {Armstrong}, {Zenteno},
  {Barkhouse}, {Bertin}, {Buckley-Geer}, {Hansen}, {High}, {Lin},
  {M{\"u}hlegger}, {Ngeow}, {Pierini}, {Pratt}, {Verdugo}, \&
  {Tucker}}]{suhada12}
{{\v S}uhada}, R., {et~al.} 2012, \aap, 537, A39

\bibitem[{{van Engelen} {et~al.}(2012){van Engelen}, {Keisler}, {Zahn}, {Aird},
  {Benson}, {Bleem}, {Carlstrom}, {Chang}, {Cho}, {Crawford}, {Crites}, {de
  Haan}, {Dobbs}, {Dudley}, {George}, {Halverson}, {Holder}, {Holzapfel},
  {Hoover}, {Hou}, {Hrubes}, {Joy}, {Knox}, {Lee}, {Leitch}, {Lueker},
  {Luong-Van}, {McMahon}, {Mehl}, {Meyer}, {Millea}, {Mohr}, {Montroy},
  {Natoli}, {Padin}, {Plagge}, {Pryke}, {Reichardt}, {Ruhl}, {Sayre},
  {Schaffer}, {Shaw}, {Shirokoff}, {Spieler}, {Staniszewski}, {Stark}, {Story},
  {Vanderlinde}, {Vieira}, \& {Williamson}}]{vanengelen12}
{van Engelen}, A., {et~al.} 2012, ArXiv e-prints, 1202.0546

\bibitem[{{Vanderlinde} {et~al.}(2010){Vanderlinde}, {Crawford}, {de Haan},
  {Dudley}, {Shaw}, {Ade}, {Aird}, {Benson}, {Bleem}, {Brodwin}, {Carlstrom},
  {Chang}, {Crites}, {Desai}, {Dobbs}, {Foley}, {George}, {Gladders}, {Hall},
  {Halverson}, {High}, {Holder}, {Holzapfel}, {Hrubes}, {Joy}, {Keisler},
  {Knox}, {Lee}, {Leitch}, {Loehr}, {Lueker}, {Marrone}, {McMahon}, {Mehl},
  {Meyer}, {Mohr}, {Montroy}, {Ngeow}, {Padin}, {Plagge}, {Pryke}, {Reichardt},
  {Rest}, {Ruel}, {Ruhl}, {Schaffer}, {Shirokoff}, {Song}, {Spieler},
  {Stalder}, {Staniszewski}, {Stark}, {Stubbs}, {van Engelen}, {Vieira},
  {Williamson}, {Yang}, {Zahn}, \& {Zenteno}}]{vanderlinde10}
{Vanderlinde}, K., {et~al.} 2010, \apj, 722, 1180

\bibitem[{{Wang} \& {Steinhardt}(1998)}]{wang98}
{Wang}, L., \& {Steinhardt}, P.~J. 1998, \apj, 508, 483

\bibitem[{{Williamson} {et~al.}(2011){Williamson}, {Benson}, {High},
  {Vanderlinde}, {Ade}, {Aird}, {Andersson}, {Armstrong}, {Ashby}, {Bautz},
  {Bazin}, {Bertin}, {Bleem}, {Bonamente}, {Brodwin}, {Carlstrom}, {Chang},
  {Chapman}, {Clocchiatti}, {Crawford}, {Crites}, {de Haan}, {Desai}, {Dobbs},
  {Dudley}, {Fazio}, {Foley}, {Forman}, {Garmire}, {George}, {Gladders},
  {Gonzalez}, {Halverson}, {Holder}, {Holzapfel}, {Hoover}, {Hrubes}, {Jones},
  {Joy}, {Keisler}, {Knox}, {Lee}, {Leitch}, {Lueker}, {Luong-Van}, {Marrone},
  {McMahon}, {Mehl}, {Meyer}, {Mohr}, {Montroy}, {Murray}, {Padin}, {Plagge},
  {Pryke}, {Reichardt}, {Rest}, {Ruel}, {Ruhl}, {Saliwanchik}, {Saro},
  {Schaffer}, {Shaw}, {Shirokoff}, {Song}, {Spieler}, {Stalder}, {Stanford},
  {Staniszewski}, {Stark}, {Story}, {Stubbs}, {Vieira}, {Vikhlinin}, \&
  {Zenteno}}]{williamson11}
{Williamson}, R., {et~al.} 2011, \apj, 738, 139

\bibitem[{{Yee} {et~al.}(2000){Yee}, {Morris}, {Lin}, {Carlberg}, {Hall},
  {Sawicki}, {Patton}, {Wirth}, {Ellingson}, \& {Shepherd}}]{Yee00}
{Yee}, H.~K.~C., {et~al.} 2000, \apjs, 129, 475

\bibitem[{{Zenteno} {et~al.}(2011){Zenteno}, {Song}, {Desai}, {Armstrong},
  {Mohr}, {Ngeow}, {Barkhouse}, {Allam}, {Andersson}, {Bazin}, {Benson},
  {Bertin}, {Brodwin}, {Buckley-Geer}, {Hansen}, {High}, {Lin}, {Lin}, {Liu},
  {Rest}, {Smith}, {Stalder}, {Stark}, {Tucker}, \& {Yang}}]{zenteno11}
{Zenteno}, A., {et~al.} 2011, \apj, 734, 3

\end{thebibliography}

\appendix

\section{BCS Catalog Description}
We created an ASCII catalog files which is obtained from the catalogs of each individual tile and after
removing duplicates. The description of each column in the BCS catalog is provided in Tab.~\ref{tab:bcscatalog}.

\begin{deluxetable}{lccc}
\tighten
\tabletypesize{\scriptsize}
\tablecaption{Details of BCS Catalogs}
\tablewidth{0pt}
\tablehead{
\colhead{Column} & \colhead{Parameter}  & \colhead{Units}  & \colhead{Definition}
}
1 & {\tt tilename} & - & Name of tile \\
2 & {\tt objectid}  & - & ID from DESDM database table {\tt coadd\_objects} \\
3 & {\tt RA}        & degree & Right Ascension \\
4 & {\tt DEC}     & degree & Declination \\
5 & {\tt mag\_model\_g} & AB mag & Model magnitude ($g$) \\
6 & {\tt magerr\_model\_g} & AB mag & Error in model magnitude ($g$) \\
7 & {\tt mag\_auto\_g} & AB mag & Kron magnitude ($g$) \\
8 & {\tt magerr\_auto\_g} & AB mag & Error in Kron magnitude ($g$) \\
9 & {\tt mag\_psf\_g} & AB mag & PSF magnitude ($g$) \\
10 & {\tt magerr\_psf\_g} & AB mag & Error in PSF magnitude ($g$) \\
11 & {\tt mag\_petro\_g} & AB mag & Petrosian magnitude ($g$) \\
12 & {\tt magerr\_petro\_g} & AB mag & Error in Petrosian magnitude ($g$) \\
13. & {\tt mag\_aper3\_g} & AB mag & Magnitude in 3-arcsec aperture ($g$) \\
14 & {\tt magerr\_aper3\_g} & AB mag & Magnitude error in 3-arcsec aperture ($g$) \\
15 & {\tt flags\_g} & - & {\sc SExtractor} flag ($g$) \\
16 & {\tt class\_star\_g} & - & {\sc SExtractor} star/galaxy separator \\ 
17 & {\tt spread\_model\_g} & - & Difference in PSF and S\'ersic magnitude ($g$) \\
18 & {\tt spread\_modelerr\_g} & - & Error in {\tt spread\_model} ($g$) \\
19 & {\tt mag\_model\_r} & AB mag & Model magnitude ($r$) \\
20 & {\tt magerr\_model\_r} & AB mag & Error in model magnitude ($r$) \\
21 & {\tt mag\_auto\_r} & AB mag & Kron magnitude ($r$) \\
22 & {\tt magerr\_auto\_r} & AB mag & Error in Kron magnitude ($r$) \\
23 & {\tt mag\_psf\_r} & AB mag & PSF magnitude ($r$) \\
24 & {\tt magerr\_psf\_r} & AB mag & Error in PSF magnitude ($r$) \\
25 & {\tt mag\_petro\_r} & AB mag & Petrosian magnitude ($r$) \\
26 & {\tt magerr\_petro\_r} & AB mag & Error in Petrosian magnitude ($r$) \\
27 & {\tt mag\_aper3\_r} & AB mag & Magnitude in 3-arcsec aperture ($r$) \\
28 & {\tt magerr\_aper3\_r} & AB mag & Magnitude error in 3-arcsec aperture ($r$) \\
29 & {\tt flags\_r} & - & {\sc SExtractor} flag ($r$) \\
30 & {\tt class\_star\_r} & - & {\sc SExtractor} star/galaxy separator \\
31 & {\tt spread\_model\_r} & - & Difference in PSF and S\'ersic magnitude ($r$) \\
32 & {\tt spread\_modelerr\_r} & - & Error in {\tt spread\_model} ($r$) \\
33 & {\tt mag\_model\_i} & AB mag & Model magnitude ($i$) \\
34 & {\tt magerr\_model\_i} & AB mag & Error in model magnitude ($i$) \\
35 & {\tt mag\_auto\_i} & AB mag & Kron magnitude ($i$) \\
36 & {\tt magerr\_auto\_i} & AB mag & Error in Kron magnitude ($i$) \\
37 & {\tt mag\_psf\_i} & AB mag & PSF magnitude ($i$) \\
38 & {\tt magerr\_psf\_i} & AB mag & Error in PSF magnitude ($i$) \\
39 & {\tt mag\_petro\_i} & AB mag & Petrosian magnitude ($i$) \\
40 & {\tt magerr\_petro\_i} & AB mag & Error in Petrosian magnitude ($i$) \\
41 & {\tt mag\_aper3\_i} & AB mag & Magnitude in 3-arcsec aperture ($i$) \\
42 & {\tt magerr\_aper3\_i} & AB mag & Magnitude error in 3-arcsec aperture ($i$) \\
43 & {\tt flags\_i} & - & {\sc SExtractor} flag ($i$) \\
44 & {\tt class\_star\_i} & - & {\sc SExtractor} star/galaxy separator \\
45 & {\tt spread\_model\_i} & - & Difference in PSF and S\'ersic magnitude ($i$) \\
46 & {\tt spread\_modelerr\_i} & - & Error in {\tt spread\_model} ($i$) \\
47 & {\tt mag\_model\_z} & AB mag & Model magnitude ($z$) \\
48 & {\tt magerr\_model\_z} & AB mag & Error in model magnitude ($z$) \\
49 & {\tt mag\_auto\_z} & AB mag & Kron magnitude ($z$) \\
50 & {\tt magerr\_auto\_z} & AB mag & Error in Kron magnitude ($z$) \\
51 & {\tt mag\_psf\_z} & AB mag & PSF magnitude ($z$) \\
52 & {\tt magerr\_psf\_z} & AB mag & Error in PSF magnitude ($z$) \\
53 & {\tt mag\_petro\_z} & AB mag & Petrosian magnitude ($z$) \\
54 & {\tt magerr\_petro\_z} & AB mag & Error in Petrosian magnitude ($z$) \\
55 & {\tt mag\_aper3\_z} & AB mag & Magnitude in 3-arcsec aperture ($z$) \\
56 & {\tt magerr\_aper3\_z} & AB mag & Magnitude error in 3-arcsec aperture ($z$) \\
57 & {\tt flags\_z} & - & {\sc SExtractor} flag ($z$) \\
58 & {\tt class\_star\_z} & - & {\sc SExtractor} star/galaxy separator \\
59 & {\tt spread\_model\_z} & - & Difference in PSF and S\'ersic magnitude ($z$) \\
60 & {\tt spread\_modelerr\_z} & - & Error in {\tt spread\_model} ($z$) \\
61 & {\tt z\_phot} & - & Photometric redshift \\
62 & {\tt z\_phot\_err} & - & Photometric redshift error \\
63 & {\tt z\_phot\_flag} & - & Within \annz\ training set parameter space

\enddata
\tablecomments{Explanation and contents of catalogs in the BCS survey release. More details on some of the parameters
can be found in the {\sc SExtractor} manual. The magnitudes are corrected for
galactic extinction.} 
\label{tab:bcscatalog}
\end{deluxetable}

\end{document}